\documentclass[12pt]{article}
\usepackage{amsmath,amsfonts,amssymb}
\usepackage{hyperref}
\usepackage{graphicx}
\usepackage{feynmp}
\usepackage[dvips]{color}
\usepackage{epsfig}
\usepackage{psfrag}
\usepackage{sectsty}
\subsubsectionfont{\it\centering}

\makeatletter\@addtoreset{equation}{section}\makeatother

\DeclareMathOperator{\Tr}{Tr}

\def\bC {\mathbb{C}}

\def\bR {\mathbb{R}}
\def\bZ {\mathbb{Z}}

\newcommand{\beq}{\begin{equation}}
\newcommand{\eeq}{\end{equation}}
\newcommand{\bal}{\begin{equation}\begin{aligned}}
\newcommand{\eal}{\end{aligned}\end{equation}}
\newcommand{\bea}{\begin{eqnarray}}
\newcommand{\eea}{\end{eqnarray}}

\newcommand{\vev}[1]{{\left< {#1} \right>}}

\newcommand{\eqn}[1]{(\ref{#1})}

\newcommand{\address}[1]{\vbox{\center\em#1}}
\renewcommand{\title}[1]{\vbox{\center\LARGE{#1}}\vspace{5mm}}

\newcommand{\cF}{{\mathcal F}}

\newcommand{\cL}{{\mathcal L}}

\newcommand{\cN}{{\mathcal N}}

\newcommand{\cO}{{\mathcal O}}

\newcommand{\ci}{{\mathcal I}}

\newcommand{\cZ}{{\mathcal Z}}
\newcommand{\al}{\alpha}
\newcommand{\gfrak}{{\mathfrak g}}
\newcommand{\su}{{\mathfrak s\mathfrak u}}
\newcommand{\hfrak}{{\mathfrak h}}
\newcommand{\LR}{{}^L\negthinspace R}
\newcommand{\LG}{{}^L\negthinspace\hspace{.4mm} G}
\newcommand{\Lgfrak}{{}^L\negthinspace\hspace{.4mm} {\mathfrak g}}
\begin{document}
\bibliographystyle{utphys}
\begin{fmffile}{graphs}

\begin{titlepage}
\begin{center}
\vspace{5mm}
\hfill {\tt HU-EP-10/07}\\
\vspace{14mm}

\title{\hskip-15pt The Virtue of Defects in 4D Gauge Theories and 2D CFTs}
\vspace{15mm}

\renewcommand{\thefootnote}{$\alph{footnote}$}

Nadav Drukker\footnote{\href{mailto:drukker@physik.hu-berlin.de}
{\tt drukker@physik.hu-berlin.de}},
Davide Gaiotto\footnote{\href{mailto:dgaiotto@ias.edu}
{\tt dgaiotto@ias.edu}},
and
Jaume Gomis\footnote{\href{mailto:jgomis@perimeterinstitute.ca}
{\tt jgomis@perimeterinstitute.ca}}
\vskip 5mm
\address{
${}^{a}$Institut f\"ur Physik, Humboldt-Universit\"at zu Berlin,\\
Newtonstra\ss e 15, D-12489 Berlin, Germany
}
\address{
${}^{b}$School of Natural Sciences, Institute for Advanced Study,\\
Princeton, NJ 08540, USA}
\address{
${}^{c}$Perimeter Institute for Theoretical Physics,\\
Waterloo, Ontario, N2L 2Y5, Canada}

\renewcommand{\thefootnote}{\arabic{footnote}}
\setcounter{footnote}{0}

\end{center}

\vspace{8mm}
\abstract{
\medskip
\normalsize{
\noindent
We advance a correspondence between the topological defect operators 
in Liouville and Toda conformal field theories -- which we construct -- 
and loop operators and domain wall operators in four dimensional 
${\cal N}=2$ supersymmetric gauge theories on $S^4$. Our computation of the   correlation functions 
in Liouville/Toda theory in the presence of topological defect
operators, which are supported on curves on the Riemann surface, yields the exact answer for the partition function of   four dimensional gauge theories in the 
presence of various walls and loop operators;   results which we can quantitatively substantiate with an independent gauge theory analysis.
As an  interesting
outcome of this work for two dimensional conformal field theories, we prove that
topological defect operators and the Verlinde loop operators are different descriptions of the  same operators.}
}
\vfill

\end{titlepage}

\tableofcontents
%%%%%%%%%%%%%%%%%%%%%%%%%%%%%%

\newpage
\section{Introduction and Summary}
\label{sec:intro}

Solving exactly four dimensional non-supersymmetric gauge theories is currently out of reach. 
Presently, we can at best compute observables in gauge theories 
in a perturbative expansion. These observables include 
the S-matrix, the correlation function of local, gauge invariant 
operators and the expectation value of non-local gauge invariant 
operators, such as Wilson loops, 't~Hooft loops and surface operators. 
Little is known about the behaviour of these observables in the strong 
coupling regime. 

The recent observation \cite{AGT} that the partition function of a class of four 
dimensional ${\cal N}=2$ gauge theories on $S^4$ as computed by Pestun 
\cite{Pestun:2007rz}  is captured by a correlation function in two dimensional 
Liouville CFT, has provided a novel arena in which to obtain new exact results 
in four dimensional gauge theories. In \cite{Alday:2009fs,Drukker:2009id} 
the exact expression for the expectation value of certain Wilson-'t~Hooft 
operators in these gauge theories were found by computing correlation functions 
in the presence of Verlinde loop operators \cite{Verlinde:1988sn} in Liouville CFT.
These connections link the study of  supersymmetric four dimensional gauge theories to the beautiful subject of two dimensional  CFTs,  and provide  a new     toolkit 
in which to study gauge theories.

In this paper we give the four dimensional gauge theory interpretation of a rather rich 
and interesting class of observables supported on curves on a Riemann surface in two 
dimensional CFTs: {\it topological defect operators}. We present exact results 
for the expectation value of various 
defect operators in four dimensional ${\cal N}=2$ supersymmetric 
gauge theories on $S^4$ from the computation of 
correlation functions in the presence of topological loop operators 
(and generalized loop operators) in Liouville/Toda CFTs.

Topological loop operators \cite{Petkova:2000ip} are labeled by a (homotopy class of a) 
curve on the Riemann surface where the CFT is defined and by a 
representation of the chiral algebra ${\cal A}$ of the CFT (the 
Virasoro algebra for Liouville theory and a $W$-algebra for 
Toda CFTs). We construct the topological loop operators in 
Liouville/Toda CFTs and show that they correspond to the 
following observables in four dimensional gauge theory:
\begin{description}
\item[$\bullet$]
Domain Wall Operators
\item[$\bullet$]
Loop Operators
\end{description}
depending on which type of representation of ${\cal A}$ 
(non-degenerate vs. degenerate) labels the associated 
topological loop operator in Liouville/Toda CFTs.

The key result that allows us to establish these correspondences relies in the 
explicit construction of the topological loop operator in Liouville/Toda CFTs.
The explicit formulae for the topological defect 
operators for Liouville/Toda CFTs, found in Section~\ref{sec:topliouville}, 
allows us to identify correlation functions with an insertion of a topological 
defect operator with the partition function of four dimensional gauge 
theories in the presence of domain wall and loop operators. In our study of 
topological defect operators  we make a  new connection with   another class
of interesting loop operators in 2d CFT, the Verlinde loop operators.  
We establish a   correspondence between these two operators by proving  that
Verlinde loop 
operators \cite{Verlinde:1988sn} are isomorphic to  topological defect operators 
\cite{Petkova:2000ip}. This result, which we prove in 
Section~\ref{sec:topdefects}, is a consequence of the associativity of 
the operator product expansion in 2d CFTs.\footnote{The computation 
of the Verlinde loop operators in Liouville theory was independently
phrased in the language of topological defects in \cite{Petkova:2009pe}.}
This observation results in a drastic simplification in the CFT calculation 
of the Verlinde loop operators, as topological defects are formulated in 
terms of more accessible data of the CFT: the modular matrices for the 
characters of the CFT. We believe that this novel connection may find interesting
applications in 2d CFTs.

We show that all classes of loop operators we construct in Liouville/Toda 
CFTs are realized as a combination of a domain wall and a Wilson loop in
at least one duality frame of the four dimensional gauge theory.%
\footnote{Corresponding to a duality frame where the topological loop 
operator wraps a tube in a pants decomposition of the Riemann surface.} 
Therefore, in other duality frames -- obtained by the action of S-duality -- 
topological loop operators in Liouville/Toda CFTs describe a combination 
of a domain wall and a Wilson-'t~Hooft loop operator in gauge theory. 
We describe in detail the gauge theory domain walls in a choice of 
duality frame (Lagrangian description), while in other duality frames 
it is defined implicitly through S-duality.

Topological loop operators in Liouville/Toda CFTs provide a framework in which to study gauge 
theory loop operators. We prove that topological defect operator labeled by 
a degenerate representation of the chiral algebra of Liouville/Toda CFTs -- 
that is a representation\footnote{This is for $b=1$, when the correspondence between 2d 
CFT and 4d gauge theory on $S^4$ holds. $b$ determines the central 
charge of the Liouville/Toda CFTs.} of the gauge 
group of the corresponding four dimensional gauge theory -- precisely 
inserts a Wilson loop operator in an arbitrary represention $R$ of the gauge group 
\beq
\hbox{Tr}_R\, e^{2\pi i a}
\eeq
into the gauge theory partition function on $S^4$ \cite{Pestun:2007rz}. 
These results clinch the Liouville/Toda CFT description of gauge theory loop 
operators in terms of topological defect operators in the CFT.

We show that certain other topological loop operators in 
Liouville/Toda CFT admit an interpretation as domain walls 
localized at the equatorial $S^3$ inside the $S^4$ 
where the four dimensional gauge theory is defined:

\noindent
$\bullet$ Janus Domain Walls: We construct 
generalized loop operators in CFTs that have the effect of changing 
the complex structure of the Riemann surface on the holomorphic 
sector with respect to the complex structure on the antiholomorphic 
sector of the CFT. Insertion of these operators define CFT correlators 
where the holomorphic and antiholomorphic sectors live on {\it different} 
Riemann surfaces. The correlation function of these generalized loop 
operators
are shown to exactly reproduce the partition function 
of the supersymmetric Janus domain walls of the corresponding 
four dimensional ${\cal N}=2$ gauge theories on $S^4$, which we 
compute using localization. These are domain 
walls across which the complexified coupling constant of the theory 
jumps, such that the coupling constant 
on the northern hemisphere of $S^4$ is different from that on the 
southern hemisphere of $S^4$. 

\noindent
$\bullet$ 
Symmetry Breaking Domain Walls: A very large family of topological defect 
operators in Liouville/Toda 
CFTs are labeled by non-degenerate and semi-degenerate representations 
of the chiral algebra ${\cal A}$. We make a rather explicit identification 
of the effect of inserting these loop operators in CFT correlators with the 
partition function of a supersymmetric domain walls 
supported on the equatorial $S^3$ of the corresponding ${\cal N}=2$ 
gauge theory. These gauge theory domain walls reduce the gauge 
group $G$ of the four dimensional gauge theory down to a subgroup 
$H$ at the location of the wall. By analyzing the effect of the insertion 
of such a domain wall to the gauge theory partition function on $S^4$ 
we exactly reproduce the formula the semi-degenerate topological 
defect operators in Liouville/Toda CFTs.

\noindent
$\bullet$ Duality Walls:
These are domain walls 
between two four dimensional gauge theories, where the theory on one 
side of the wall is the S-dual of the other. These walls have the property 
that operators transported across them are acted on by the action of 
S-duality. We propose that these duality walls
correspond to   Liouville/Toda CFT correlation functions enriched 
by the action of an element of the Moore-Seiberg groupoid ${\cal G}$ \cite{Moore:1988qv}
(which corresponds to the S-duality group of the corresponding four dimensional 
gauge theory). This correspondence predicts that the partition function of 
the three dimensional ${\cal N}=2$ superconformal field theory on the 
equatorial $S^3$ is captured by the integral kernel that implements the 
transformation on the CFT conformal blocks of the associated Moore-Seiberg 
groupoid element.

In summary, we have found a rather complete realization of topological 
defect operators in Liouville/Toda CFTs in terms of defect operators 
(loop and domain wall operators) in four dimensional ${\cal N}=2$ 
supersymmetric gauge theories on $S^4$. Our computations in 
Liouville/Toda CFTs allow for the exact computation of these interesting 
observables in four dimensional gauge theories.

The plan of the rest of the paper is as follows. In Section~\ref{sec:topdefects} 
we prove that Verlinde loop operators are isomorphic to topological loop 
operators in a general CFT using the Moore-Seiberg relations of 2d CFT. 
Section~\ref{sec:topliouville} contains the explicit construction of topological 
loop operators in Liouville theory and arbitrary Toda CFTs as well as 
generalized loop operators, which change the complex structure of the 
Riemann surface. This entails determining the characters of these CFTs 
for the various representations as well as computing the associated 
modular matrices. In Section~\ref{sec:ZZ-loops} it is shown 
that topological defect operators based on degenerate representations 
exactly capture the insertion of a Wilson loop operator in the partition 
function of the four dimensional gauge theory on $S^4$. Section~\ref{pestun} 
shows that the other topological defects in Liouville/Toda CFTs admit an 
interpretation as domain wall operators in the four dimensional gauge 
theory. In section~\ref{sec:M} we use the construction of the four 
dimensional gauge theories using M5-branes to shed further light 
on the identification between topological defects and gauge theory 
defects presented in the previous sections. Some extra calculations are given in appendices. In Appendix~\ref{sec:susy} 
we present the different subgroups of the 4d $\cN=2$ supergroup that 
are preserved by $1/2$ BPS domain walls, surface operators, loop operators and 
local operators. In Appendix~\ref{sec:glue-paste} we present some more 
information on the construction of boundaries and domain walls in 
4d $\cN=2$ gauge theories and in particular the Janus domain wall. 
Appendix~\ref{sec:S-duality} discusses the duality wall associated to 
the S-duality of $\cN=4$ SYM.

\section{Topological Defects and Loop Operators in 2D CFT}
\label{sec:topdefects}

In this section we establish a novel correspondence between two 
seemingly unrelated loop operators in 2d CFT -- known as topological 
defect operators and Verlinde loop operators -- by proving that they 
are isomorphic. This new connection is of interest and relevance for 
2d CFTs and does not rely on their physical interpretation in 4d 
${\cal N}=2$ gauge theories, which is the main focus of the rest 
of the paper. This section and Section~\ref{sec:topliouville} 
are self-contained and may be read independently.

\subsection{Topological Defect Operators}

Topological defect operators \cite{Petkova:2000ip} 
(see also e.g. \cite{Petkova:2001ag,Fuchs:2002cm,Bachas:2004sy}) 
in a 2d CFT are operators supported on closed curves
characterized by the property that they commute with {\it both} the
holomorphic and antiholomorphic energy-momentum tensor of the CFT.%
\footnote{More generally, topological defect operators can glue two {\it different} 
CFTs with the same central charge \cite{Bachas:2001vj}. In this paper we focus our attention to the 
case where the same CFT appears on both sides of the defect.}
Since the energy-momentum tensor generates arbitrary conformal transformations,
which can deform the curve, the geometric characterization of topological 
defect operators is in terms of homotopy classes of non-intersecting curves on the punctured 
Riemann surface where the CFT is defined.%
\footnote{A important natural generalization are topological 
defect webs, which we will consider only briefly.}
In formulae, a topological defect 
operator supported on an (oriented) curve $p$, which we denote by 
${\cal O}_\mu(p)$, is defined by
\beq
[T(z), {\cal O}_\mu(p)]=0,\qquad\qquad [\overline{T}(\bar{z}), {\cal O}_\mu(p)]=0\,,
\label{topdefect}
\eeq
where $T$ and $\overline{T}$ denote the holomorphic and antiholomorphic
energy-momentum tensors of the CFT, which generate the two copies of the
Virasoro algebra ${\it Vir}\otimes {\it Vir}$ preserved by the topological defect 
operator. The label $\mu$ in ${\cal O}_\mu(p)$ specifies the quantum number 
(i.e. the representation) of the operator.

\begin{figure}[t]
\centering
\epsfig{file=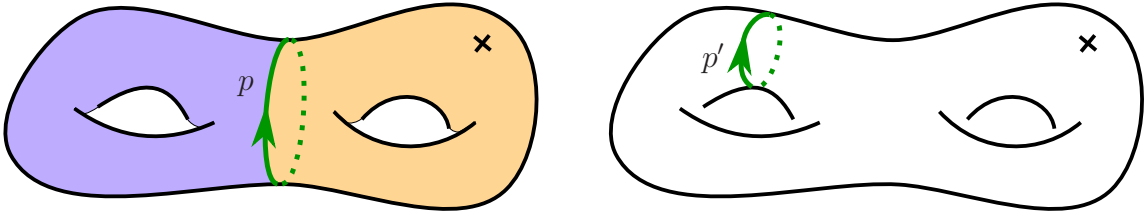,width=6in
\psfrag{c}{$p$}
\psfrag{c'}{$p'$}}
\parbox{5in}{\caption{Two curves on a genus-2 Riemann surface with one puncture. 
The curve $p$ separates the surface into two regions on which two 
distinct CFTs can be defined, joined by a topological defect. The curve 
$p'$ does not split the surface into two disconnected surfaces, but rather to 
a single surface with two boundaries. One can   define a topological defect 
identifying two copies of the same CFT along $p'$.}
\label{fig:top}}
\end{figure}

\begin{figure}[t]
\centering
$$
\begin{tabular}{cp{.2in}c}
\epsfig{file=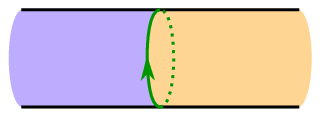,height=.8in}&&
\epsfig{file=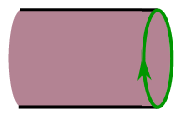,height=.8in}\\[3mm]
$\text{CFT}_1$\hskip.7in $\text{CFT}_2$&&
$\text{CFT}_1\otimes\overline{\text{CFT}}_2$
\end{tabular}
$$
\parbox{5in}{\caption{Locally a topological defect can always be represented as 
separating two CFTs on the cylinder. This is equivalent to a boundary condition for the 
tensor product of the two CFTs.}
\label{fig:folding}}
\end{figure}

Topological defect operators in a CFT can be mapped to conformally invariant boundary
conditions (BCFTs) of the doubled theory $\text{CFT}\otimes\overline{\text{CFT}}$ using
the folding trick, whereby the CFT on one side of the defect is mapped to the
other one (see Figure~\ref{fig:folding}).\footnote{The bar on the 
folded copy of the CFT reflects that unfolding
acts by parity and exchanges the holomorphic and antiholomorphic sectors of
the folded CFT.}
Nontrivial topological defect operators correspond to the so-called permutation
BCFTs, defined by gluing the stress-energy tensor of CFT with that of
$\overline{\text{CFT}}$ on the boundary. BCFTs which are a direct product of a 
BCFT of CFT and $\overline{\text{CFT}}$ yield trivial topological defect operators 
upon folding, as the CFTs at the two sides of the defect
are completely decoupled. The construction of topological defects is therefore
tantamount to the construction of permutation BCFTs of
$\text{CFT}\otimes\overline{\text{CFT}}$.

Rational CFTs (RCFTs) provide a special class of theories for which 
topological defects can be completely understood. The holomorphic 
data of a CFT depends on the choice of a chiral algebra ${\cal A}$, 
which necessarily contains the Virasoro algebra ${\it Vir}$ in its 
enveloping algebra $U({\cal A})$.\footnote{For instance, when 
${\cal A}$ is an affine algebra, the Virasoro generators are 
quadratic forms in the currents generating ${\cal A}$ and 
${\it Vir}\subset U({\cal A})$, while for W-algebras 
${\it Vir}\subset {\cal A}$.} A RCFT is characterized by having a finite 
set $\ci$ of irreducible representations ${\cal V}_\alpha$ of the algebra 
${\cal A}$, with associated characters $\chi_\alpha(q)$, where $\alpha\in \ci$. 
Topological defect operators invariant under
${\cal A}\otimes{\cal A}$ can be classified in RCFTs and are determined by the equations
\beq
\big[W(z), {\cal O}_\mu(p)\big]=0\,,
\qquad
\big[\overline{W}(\bar{z}), {\cal O}_\mu(p)\big]=0\,,
\label{topdefectchiral}
\eeq
where the operators $W(z)$ and $\overline{W}({\bar z})$ generate
${\cal A} \otimes {\cal A}$.

For diagonal RCFTs, defined by the torus partition function
\beq
Z=\sum_{\alpha\in {\cal I}} \chi_\alpha(q) {\chi_{\alpha^*}(\bar{q})}\,,
\eeq
where $\alpha^*$ labels the representation conjugate\footnote{The 
product of a representation $\alpha$ with its conjugate $\alpha^*$ 
contains the identity (vacuum) representation $1$, i.e. 
${\cal V}_\alpha\otimes {\cal V}_{\alpha^*} \supset {\cal V}_1$. They are 
the unique pair of representations for which the fusion coefficient 
${\cal N}_{\alpha\alpha^*}^{~~1}=1$.} to ${\cal V}_\alpha$,
the number of linearly independent topological defect operators
${\cal O}_\mu(p)$ is given by the cardinality $|\ci|$ of the representation
set $\ci$:\footnote{For a general RCFT, defined
by the partition function
$Z=\sum_{\alpha,\bar{\alpha}\in I} 
Z_{\alpha\bar{\alpha}}\chi_\alpha(q){\chi_{{\bar \alpha}}(\bar{q})}$, 
where $Z_{\alpha\bar{\alpha}}\in \bZ_+$, the
number of linearly independent topological defect operators is given by the
cardinality $|{\cal E}|$ of the set ${\cal E}=\{\alpha| Z_{\alpha\alpha^*}\neq 0\}$, known as the 
set of {\it exponents}.}
\beq
\{{\cal O}_\mu(p)\,|~\mu\in\ci\}\,.
\eeq

In the doubled theory the commutation relations \eqn{topdefectchiral} 
become the equations defining the permutation Ishibashi states. Indeed, 
in radial quantization the equation reads%
\footnote{In this formula the overline denotes the antiholomorphic sector.}
\beq
\left(W^{{(s)}1}_{n}-(-1)^{s} \overline{W}_{-n}^{{(s)}2}\right)|\alpha,\alpha^*\rangle\!\rangle
=
\left(W^{{(s)}2}_{n}-(-1)^{s}\overline{W}_{-n}^{{(s)}1}\right)|\alpha,\alpha^*\rangle\!\rangle=0\,,
\eeq
where $s$ is the spin of the generator and the superscript $1,2$ refers to the two 
copies of the CFT. A complete basis of permutation Ishibashi states is given by 
\beq
|\alpha,\alpha^*\rangle\!\rangle
\equiv\sum_{\{k\},\{l\}} |\alpha,k;\alpha^*,l \rangle\otimes |\alpha,l;\alpha^*,k \rangle\,,
\eeq
with $\alpha\in\ci$ any of the representations of $\cal A$ and $ \sum_{\{k\},\{l\}}$ 
is the sum over all orthonormalized descendant states\footnote{If the states
are not orthonormalized the inverse matrix of inner products should be inserted into the expression.}
in the representation 
${\cal V}_\alpha\otimes\overline{{\cal V}}_{\alpha^*}$ of ${\cal A}$ .

The overlap of permutation 
Ishibashi states yield the product of the holomorphic and antiholomorphic characters
\beq
\langle\!\langle \mu,\mu^*| q^{L_0-c/24} \,
\overline{q}^{\overline{L}_0-c/24}|\alpha,\alpha^*\rangle\!\rangle
= \delta_{\mu\alpha} \chi_\alpha(q)\chi_{\alpha^*}(\bar{q})\,.
\eeq

The fact that
$\cO_\mu(p)$ commutes with both copies of the chiral algebra
${\cal A}$ in (\ref{topdefectchiral}) implies that a topological defect operator is
proportional to the identify operator in each subspace
${\cal V}_\alpha\otimes \overline{{\cal V}}_{\alpha^*}$ of the Hilbert space. Therefore
${\cal O}_\mu(p)$ must be a sum of projectors 
\beq
{\cal O}_\mu(p)=\sum_{\alpha\in \ci} D_{\mu \alpha} 
\sum_{\{k\},\{l\}} |\alpha,k;\alpha^*,l \rangle\otimes \langle \alpha,k;\alpha^*,l|\,,
\label{projectors}
\eeq

The coefficients $D_{\mu \alpha}$ are determined by demanding that the partition 
function on the torus in the presence of a topological defect admits a 
consistent Hilbert space interpretation, akin to the familiar Cardy condition 
in the context of BCFT. Consistency requires that%
\footnote{In contrast, the 
expression for the Cardy boundary states in RCFT is given by 
$|\mu\rangle=\sum_{\alpha\in \ci}{S_{\mu \alpha}\over \sqrt{S_{1\alpha}}}|\alpha\rangle$. 
The change of power in the denominator is due to the fact that, 
as already mentioned, a topological defect in a CFT is directly 
related to a boundary state for the doubled CFT.}
\cite{Petkova:2000ip}
\beq
D_{\mu \alpha}={S_{\mu \alpha}\over S_{1\alpha}}\,,
\label{modular}
\eeq
where $S_{\mu\alpha}$ is the modular matrix of the CFT, defined by the way 
the characters of ${\cal A}$ behave under a modular transformation
\beq
\chi_\mu(q)=\sum_{\alpha\in I} S_{\mu\alpha} \chi_\alpha(\tilde{q})\,,
\eeq
where $q=\exp(2\pi i \tau)$ and $\tilde{q}=\exp(-2\pi i /\tau)$, and where 
$1\in \ci$ denotes the trivial (vacuum) representation of the chiral algebra 
${\cal A}$. Therefore, we find that topological loop operators are labeled 
by a representation $\mu\in {\cal I}$ of the chiral algebra ${\cal A}$.

Reversing the orientation of the curve $p$ in $C_{g,n}$ changes the 
representation label of the topological line operator to the conjugate 
representation, that is ${\cal O}_{\mu}(p^{-1})={\cal O}_{\mu^*}(p)$. 
It is therefore important to keep track of the orientation of curves 
supporting topological defect operators in CFTs based on chiral 
algebras ${\cal A}$ admitting complex representations.

Even though no general classification of topological defects in non-rational CFTs
or in RCFTs preserving ${\it Vir}\otimes{\it Vir}$ is available,
some explicit constructions are known. We will later show that topological
defect operators in some non-rational CFTs -- such as Liouville and Toda theories -- share
similarities with the corresponding operators in RCFTs.

Topological defect operators can be used to construct observables that 
go beyond the usual CFT correlation functions, thereby enriching the 
familiar correlation functions of local operators. On a genus $g$ Riemann 
surface with $n$ punctures, which we denote by $C_{g,n}$, topological 
defect operators supported on curves in $C_{g,n}$ 
can be inserted.
For instance, 
one can define the expectation value of the topological defect 
operator ${\cal O}_\mu(p)$ in the presence of the $n$ local operators
\begin{equation}
\label{toploop}
\big\langle {\cal O}_\mu(p)\big\rangle_{C_{g,n}}^{}\equiv 
\Big<{\cal O}_\mu(p)\prod_{i=1}^n V_{m_i} \Big>_{C_{g,n}}\,.
\end{equation}

Topological defect operators have important physical implications and have 
provided insights into symmetries, duality transformations, RG flows and 
boundary states in CFTs (see e.g. 
\cite{Quella:2002ct,Graham:2003nc,Frohlich:2004ef,Frohlich:2006ch,
Quella:2006de,Runkel:2007wd,Bachas:2007td}).

\begin{figure}[t]
\centering
\epsfig{file=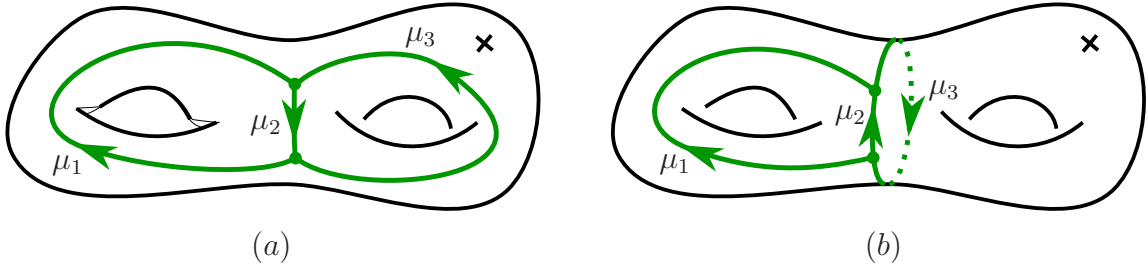,width=6in
\psfrag{m1}{$\mu_1$}
\psfrag{m2}{$\mu_2$}
\psfrag{m3}{$\mu_3$}}
$(a)\hskip3in(b)$
\parbox{5in}{\caption{Two examples of topological defect webs on a genus 
two surface with one puncture. Both have a pair of topological defect junctions. 
In $(a)$ they combine the defects with representations $\mu_1$, $\mu_2^*$, $\mu_3$ 
and $\mu_1^*$, $\mu_2$, $\mu_3^*$. In $(b)$ they combine 
$\mu_1$, $\mu_2$, $\mu_3^*$ and $\mu_1^*$, $\mu_2^*$, $\mu_3$.}
\label{fig:web}}
\end{figure}

There is an interesting generalization of the correlation function (\ref{toploop}) involving 
topological webs of line operators. Indeed three topological 
line operators can meet at a topological junction 
whose support can also be arbitrarily deformed in the 
Riemann surface without changing the value for the correlator 
(see Figure~\ref{fig:web}). Topological 
junctions joining more than three topological line operators can be split 
into products of trivalent junctions. Therefore, an arbitrary web of topological 
line operators on a Riemann surface can be generated by a ``lego" construction 
connecting oriented topological lines at trivalent topological junctions.

The number of topological junctions joining incoming topological 
line operators labeled by representations $\mu_1,\mu_2$ and $\mu_3$ is given 
by ${\cal N}_{\mu_1\mu_2}^{~~\mu_3^*}$, the fusion coefficients of the chiral 
algebra ${\cal A}$. It is also possible for a line operator with label $\mu$ to end 
on a vertex operator $V_{i,\bar j}$ if the representation $\bar j^*$ appears in the fusion 
of $\mu$ and $i$. We will denote by ${\cal O}$ a generic topological web operator,
made of strands of line operators ending either at triple intersections or on 
appropriate vertex operators.

\subsection{Conformal Blocks and Topological Loop Operators}

An arbitrary CFT correlation function can be constructed by combining
the three-point function of primary fields on the sphere with the so-called conformal
blocks. The correlator can be reduced to a product of three-point functions by
appropriately inserting complete sets of intermediate states, which includes a
sum over the primary fields of the CFT as well as the descendants. The contribution
to the correlator of the descendants, which factorizes into a holomorphic and
antiholomorphic contribution, is captured by the conformal blocks, and what 
is left is a sum over the primary labels and, possibly, over the different ways 
of fusing the primaries inside each three point function. 

\begin{figure}[t]
\centering
$$
\begin{tabular}{cp{.2in}c}
\epsfig{file=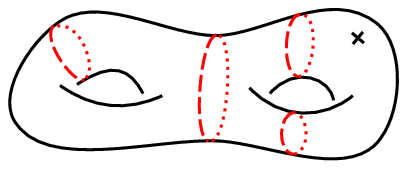,width=2.5in}&&
\epsfig{file=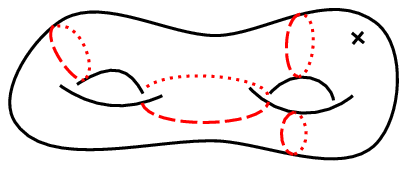,width=2.5in}\\[3mm]
\raisebox{2mm}[0mm][0mm]{
\epsfig{file=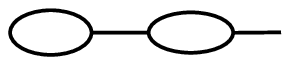,width=2in}}&&
\epsfig{file=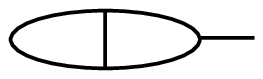,width=2in}
\end{tabular}
$$
\parbox{5in}{\caption{Two different pants decomposition of the surface $C_{2,1}$ of genus 2 with 
one puncture and the two associated trivalent graphs.}
\label{fig:pants}}
\end{figure}

Associativity of the operator product expansion (OPE) in the CFT implies that
the decomposition of a correlator into three-point functions and conformal 
blocks is not unique. Each decomposition of a correlator in $C_{g,n}$ can 
be associated with a trivalent graph $\Gamma_\sigma$, corresponding to 
a choice of sewing $\sigma$ of the Riemann surface $C_{g,n}$ from 
$2g-2+n$ pairs of pants (trinions) and $3g-3+n$ tubes (see Figure~\ref{fig:pants}). 
The sewing $\sigma$ of $C_{g,n}$ defines the conformal block
\beq
{\cal F}_{\alpha,E}^{(\sigma)}\,,
\label{BPZblock}
\eeq
The conformal block carries three types of labels: representations on the external edges,
representations on the internal edges and the choice of fusion channel at each 
trivalent vertex, which runs from $1$ to ${\cal N}_{\mu_1\mu_2}^{~~\mu_3^*}$ if 
$\mu_i$ are the labels on the three incoming edges. We use the notation 
$\alpha\equiv(\alpha_1,\ldots,\alpha_{3g-3+n},u_1, \ldots,u_{2g-2+n})$: $\alpha_i$ 
are the labels
of the representations associated with the internal edges of $\Gamma_\sigma$ and 
$u_i$ label the fusion channels. Finally, $E\equiv(m_1,\ldots,m_{n})$ label the 
representations of the external edges.

Therefore, a CFT correlator in $C_{g,n}$ admits the following holomorphically 
factorized representation\footnote{In the case of RCFTs, the integral over 
intermediate states reduces to a sum.}
\beq
\Big<\prod_{i=1}^n V_{m_i}\Big>_{C_{g,n}}=\int d\nu(\alpha)\,
{\overline \cF^{(\sigma)}_{\alpha,E}}\,\cF^{(\sigma)}_{\alpha,E}\,,
\label{correla-F}
\eeq
where the measure factor $\nu(\alpha)$ includes the product
of three-point functions corresponding to each vertex in $\Gamma_\sigma$. 
The integral is over the quantum numbers labeling the internal edges and 
trivalent vertices of $\Gamma_\sigma$, which capture the
sum over intermediate states and their selection rules.

Associativity of the OPE implies that the correlator is the same for any two 
choices of trivalent graphs $\Gamma_\sigma$ and $\Gamma_{\sigma'}$ 
of $C_{g,n}$. Since any two trivalent graphs are mapped into each other 
by an element of the so-called Moore-Seiberg groupoid ${\cal G}$ 
\cite{Moore:1988qv} (see \cite{Moore:1989vd} for a review), the 
CFT correlators are also invariant under ${\cal G}$, which includes the 
mapping class group of $C_{g,n}$ 
as an important subgroup.

This general construction applies in the presence of topological loop operators or 
webs as well. The main difference is that we will generally need to cut tubes through which 
topological lines run. As the generators of the chiral algebra ${\cal A}$ commute 
through the topological lines, the states propagating in the tubes can be still 
decomposed in irreducible representations of the chiral and antichiral algebras. 
Hence the correlation function still factorizes in holomorphic and antiholomorphic 
conformal blocks
\beq
\Big< {\cal O}\prod_{i=1}^n V_{m_i}\Big>_{C_{g,n}}=\int d\nu(\alpha')\, d\nu(\alpha)\,
{\overline \cF^{(\sigma)}_{\alpha',E}}\,{\cal O}(\alpha',\alpha) \,\cF^{(\sigma)}_{\alpha,E}\,,
\label{correla-FO}
\eeq
where ${\cal O}$ is an arbitrary topological loop or web operator.

The kernel ${\cal O}(\alpha',\alpha)$ can be computed from the product of 
appropriate three point functions in the CFT. From this expression, it is clear 
that we can interpret ${\cal O}$ as an operator {\it acting} on the space of 
conformal blocks:
\beq
\left[{\cal O}\circ \cF^{(\sigma)}\right]_{\alpha,E}=\int d\nu(\alpha')\,
{\cal O}(\alpha,\alpha') \,\cF^{(\sigma)}_{\alpha',E}\,.
\label{oper-O}
\eeq

Given two topological line operators or webs ${\cal O}$ and ${\cal O}'$ we can ``compose'' them, 
by concatenating the action on conformal blocks: 
\beq
\left[ {\cal O}\circ {\cal O}' \right] \circ \cF^{(\sigma)}={\cal O}\circ \left[ {\cal O}' \circ \cF^{(\sigma)}\right]\label{oper-O-prod}
\eeq

The composition $\left[ {\cal O}\circ {\cal O}' \right]$ can be rewritten as a sum 
over topological webs, or line operators! If the strands of ${\cal O}$ and ${\cal O}'$ 
are disjoint, the composition simply corresponds to inserting both ${\cal O}$ and 
${\cal O}'$ in the correlation function, but 
if the strands of ${\cal O}$ and ${\cal O}'$ intersect at some points $z_i$ on the 
Riemann surface, the composition is rather more interesting. We will discuss it 
briefly at the end of Section~\ref{comp}.
In particular, these topological line and web operators form an interesting algebra, 
and provide a link between conformal field theory and other mathematical 
constructions, such as quantization of Teichm\"uller space and its higher rank 
generalizations.

\subsection{Verlinde Loop Operators}
While CFT correlators are single-valued functions of the positions of the vertex operators,
conformal blocks $\cF^{(\sigma)}_{\alpha,E}$ are multivalued functions.%
\footnote{More precisely, they are sections of a projectively flat vector bundle
over the moduli space of the Riemann surface $C_{g,n}$ \cite{Friedan:1986ua}.}
The conformal blocks $\cF^{(\sigma')}_{\alpha',E}$ 
associated to $\Gamma_{\sigma'}$ are 
related to $\cF^{(\sigma)}_{\alpha,E}$ by analytic continuation.
The existence of monodromies in conformal blocks was first exploited by
E.~Verlinde to define an interesting class of observables in CFT that are supported
on homotopy classes of curves on the Riemann surface $C_{g,n}$ \cite{Verlinde:1988sn}.

The basic idea is to calculate the monodromy acquired by a chiral operator 
$V_\mu(z)$ as it circumnavigates a path $p$ in $C_{g,n}$. 
$V_\mu(z)$ can be any chiral operator in the CFT.\footnote{For nonrational theories, 
such as Liouville or Toda CFTs, loop operators can be defined for any operator in the CFT, 
even if it does not give rise to delta function normalizable states.} 
The monodromy
is calculated by enriching the conformal block $\cF^{(\sigma)}_{\alpha,E}$ associated
with the trivalent graph $\Gamma_\sigma$ by adding an external edge corresponding
to the trivial (vacuum) representation of the chiral algebra ${\cal A}$. This does not
change the value of the conformal block. In order to define the monodromy of
$V_\mu(z)$ we rewrite the identity operator as the projection to the trivial representation
of the OPE of the chiral operator $V_\mu$ with its conjugate $V_{\mu^*}$. This has the
effect of adding a trivalent vertex
to the original trivalent graph $\Gamma_\sigma$ where two of the new edges are external
and are labeled by the quantum numbers $\mu$ and $\mu^*$ while the third new one, which
carries the trivial representation $1$, is attached to
$\Gamma_\sigma$. We denote the enriched trivalent graph
$\Gamma_{\hat{\sigma}}$.

\begin{figure}[t]
\centering
$$
\begin{tabular}{cp{.2in}c}
\epsfig{file=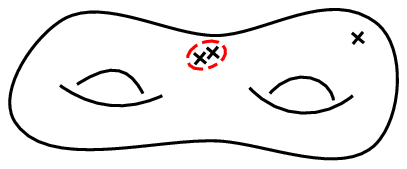,width=2.5in}&&
\epsfig{file=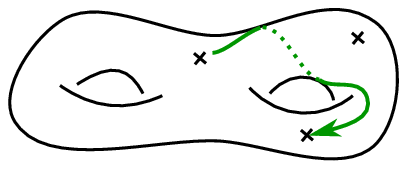,width=2.5in}\\[-2mm]
$(a)$&&$(b)$\\[3mm]
\epsfig{file=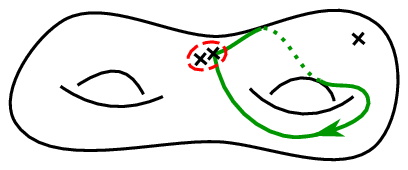,width=2.5in}&&
\epsfig{file=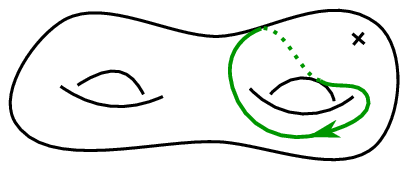,width=2.5in}\\[-2mm]
$(c)$&&$(d)$
\end{tabular}
$$
\parbox{5in}{\caption{The construction of the Verlinde loop operator. $(a)$ Two extra 
punctures carrying conjugate representations are inserted such that the channel marked by 
the red dashed line carries the identity representation. $(b)$ One of the operators transverses 
the surface along a specified path. $(c)$ When it returns to its original position we 
again project on to the identity representation in the channel separating the two punctures from 
the rest of the surface. $(d)$ If the punctures are removed we are left with the original 
surface with only the ``memory'' of the path and a representation, which label the Verlinde loop operator.}
\label{fig:Verlinde}}
\end{figure}

Given a path $p$ in $C_{g,n}$, the loop operator
${\cal L}_\mu(p)$ is defined as the monodromy acquired by $V_\mu(z)$ as it moves
around the projection of $p$ on the trivalent graph $\Gamma_\sigma$.
These loop operators are represented in terms of their action on conformal blocks
\begin{equation}
\begin{aligned}
{\cal F}^{(\sigma)}_{\alpha,E}&\longrightarrow
[{\cal L}_\mu(p)\cdot {\cal F}^{(\sigma)}]_{\alpha,E}\,,\\
{\overline {\cal F}^{(\sigma)}_{\alpha,E}}&\longrightarrow
{\overline {\cal F}^{(\sigma)}_{\alpha,E}}\,.
\label{actionmon}
\end{aligned}
\end{equation}

Transporting the chiral field $V_\mu(z)$ around the trivalent graph 
$\Gamma_\sigma$ involves intermediate steps with different 
trivalent graphs $\Gamma_{{\hat\sigma'}}$, where $V_\mu(z)$ is 
connected to different edges in $\Gamma_\sigma$. Calculating 
the monodromy requires relating the conformal blocks associated 
to different trivalent graphs. These are related by analytic continuation. 
The Moore-Seiberg groupoid ${\cal G}$, whose elements relate any 
two trivalent graphs associated with a given Riemann surface, also 
relate the conformal blocks corresponding to different trivalent graphs. 
By concatenating the moves that span the motion of $V_\mu(z)$ along 
a closed path in $\Gamma_\sigma$ the monodromy attained by 
circumnavigating along $p$ can be computed. Since the 
monodromy is invariant under continuous deformations of the 
curve $p$ on the Riemann surface, Verlinde loop operators 
${\cal L}_\mu(p)$ only depend on the homotopy class of $p$.

Just as in the case with topological defect operators, these loop 
operators can be used to define novel observables in CFT. For 
instance, the expectation of a loop operator ${\cal L}_\mu(p)$ 
in the presence of $n$ local operators is given by
\begin{equation}
\label{liouloop}
\big\langle {\cal L}_\mu(p)\big\rangle_{C_{g,n}}^{}
\,=\,\int d\nu(\alpha)\,\overline{{\cal F}}^{(\sigma)}_{\alpha,E}\,
[{\cal L}_\mu(p)\cdot {\cal F}^{(\sigma)}]_{\alpha,E}\,.
\end{equation}
Just like the usual CFT correlation functions, correlators involving loop 
operators ${\cal L}_\mu(p)$ are invariant under a change of pants 
decomposition of the Riemann surface $C_{g,n}$, and provide an 
interesting class of observables in CFTs.

\subsection{Topological Defect Operators = Verlinde Loop Operators}\label{comp}

Both topological defect operators and Verlinde loop operators constitute 
an interesting class of CFT operators supported on
curves on $C_{g,n}$ and are labeled by the homotopy class of the curve $p$ and 
a representation $\mu$ of the chiral algebra ${\cal A}$. We now prove that when 
defined along non-intersecting curves, these two seemingly 
different objects, in fact define the same loop operator.%
\footnote{Our definition of the topological loop operators requires them to 
have no self intersections. The Verlinde loop operators can be defined also along 
self-intersecting paths. There should be a generalization of this isomorphism 
to include topological defect webs and more general Verlinde loop operators.}

Let's consider ${\cal O}_\mu(p)$ and ${\cal L}_\mu(p)$ supported on a curve 
$p$ of $C_{g,n}$. In order to prove their equivalence it is sufficient to analyze the
action of these operators on correlation functions of local operators for a 
specific choice of pants decomposition of the punctured Riemann surface, 
given that these operators transform in the same way under the Moore-Seiberg groupoid.
Assuming that the curve $p$ has no self-intersections, it is always possible to choose
a pants decomposition $\sigma$ of $C_{g,n}$ using $p$ as one of the curves along
which the surface is sewn. In this pants decomposition, the path $p$
corresponds to a curve encircling one of the internal edges of the corresponding trivalent
graph $\Gamma_\sigma$.

By abuse of notation we denote the quantum number labeling this internal edge in 
$\Gamma_\sigma$ by $\alpha$. Then the topological defect 
operator ${\cal O}_\mu(p)$, described by eqns. 
(\ref{projectors})(\ref{modular}), inserts into the correlation function
\beq
{S_{\mu \alpha}\over S_{1\alpha}}\,,
\label{inserttopo}
\eeq
where $\alpha$ is a quantum number that must be summed over to 
obtain the complete correlator
\begin{equation}
\big\langle {\cal O}_\mu(p)\big\rangle_{C_{g,n}}^{}
\,=\,\int d\nu(\alpha)\,{S_{\mu \alpha}\over S_{1\alpha}}\, \overline{{\cal F}}^{(\sigma)}_{\alpha,E}\,
{\cal F}^{(\sigma)}_{\alpha,E}\,.
\end{equation}

We now want to determine the effect of the Verlinde loop operator 
${\cal L}_\mu(p)$ on the same correlation function.
As explained earlier, this is encoded by the monodromy of the chiral 
operator $V_\mu(z)$ as it circumnavigates the path $p$. 
Since this loop wraps around a single internal edge of $\Gamma_\sigma$, 
we can focus our attention on this edge, as the rest of the trivalent
graph $\Gamma_\sigma$ is unaffected by the operator ${\cal L}_\mu(p)$.

In order to evaluate the monodromy, we must determine how the 
conformal blocks based on the inequivalent trivalent graphs 
generated by the motion of $V_\mu(z)$ around $p$ are 
related. As mentioned earlier, elements in the Moore-Seiberg 
groupoid ${\cal G}$ map any two such trivalent graphs, while 
relating the corresponding conformal blocks by analytic continuation. 
It is an important result of CFT that the groupoid ${\cal G}$ is generated 
by three basic moves:%
\footnote{These are sometimes referred to as the $A$, $B$ and $S$ moves respectively. 
A combination of these operators will also realize the $s$-channel to $u$-channel 
transformation, which is also often called braiding.}
\begin{itemize}
\item
Fusion move
\item
Braiding move
\item 
S-move.
\end{itemize}

\begin{figure}[t]
\centering
$$
\text{\parbox{2in}{\epsfig{file=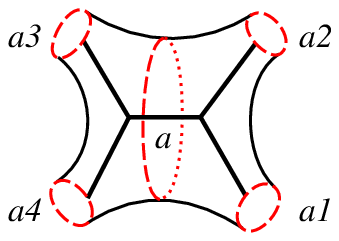,width=2in
\psfrag{a1}{$\alpha_1$}
\psfrag{a2}{$\alpha_2$}
\psfrag{a3}{$\alpha_3$}
\psfrag{a4}{$\alpha_4$}
\psfrag{a}{$\alpha$}}}}
\ \longrightarrow
\ \text{\parbox{2in}{\epsfig{file=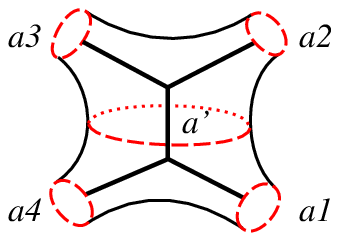,width=2in
\psfrag{a1}{$\alpha_1$}
\psfrag{a2}{$\alpha_2$}
\psfrag{a3}{$\alpha_3$}
\psfrag{a4}{$\alpha_4$}
\psfrag{a'}{$\alpha'$}}}}
$$
\parbox{5in}{\caption{The fusion move exchanges s-channel and t-channel 
graphs associated to two different pants decompositions of the 4-punctured 
sphere.}
\label{fig:F-move}}
\end{figure}

Each move acts locally on the graph and affects only the edges involved by the 
move, while the rest of edges remain untouched. We will suppress the $u_i$ labels in 
our pictures, which specify the choice of fusion channel at each vertex. The fusion move relates the 
conformal blocks associated to the following two trivalent graphs
\beq
\label{genfus}
\raisebox{0mm}[10mm][12mm]{\parbox{25mm}{\begin{center}
\begin{fmfgraph*}(20,15)
\fmfbottom{bl,br}
\fmftop{tl,tr}
\fmf{plain,width=2}{bl,b1,b2,b3,br}
\fmf{phantom}{tl,t1,t2,t3,tr}
\fmffreeze
\fmf{plain,width=2}{t1,b1}
\fmf{plain,width=2}{t3,b3}
\fmfv{label=$\alpha_4$,label.angle=-90,label.dist=8}{bl}
\fmfv{label=$\alpha_1$,label.angle=-90,label.dist=8}{br}
\fmfv{label=$\alpha$,label.angle=-90,label.dist=8}{b2}
\fmfv{label=$\alpha_3$,label.angle=180}{t1}
\fmfv{label=$\alpha_2$,label.angle=0}{t3}
\end{fmfgraph*}\end{center}}}
\equiv
\int d\al'\;
F_{\al\al'}^{}\big[\begin{smallmatrix} \al_3 & \al_2\\ \al_4 & \al_1\end{smallmatrix}\big]
\raisebox{0mm}[10mm][12mm]{\parbox{24mm}{\begin{center}
\begin{fmfgraph*}(15,15)
\fmfbottom{bl,br}
\fmftop{tl,tr}
\fmf{plain,width=2}{bl,b1,b2,b3,br}
\fmf{phantom}{tl,t1,tt1,t2,tt3,t3,tr}
\fmffreeze
\fmf{plain,width=2}{t1,c2,t3}
\fmf{plain,width=2,tension=2}{c2,b2}
\fmfv{label=$\alpha_4$,label.angle=-90,label.dist=8}{bl}
\fmfv{label=$\alpha_1$,label.angle=-90,label.dist=8}{br}
\fmfv{label=$\alpha'$,label.angle=-30}{c2}
\fmfv{label=$\alpha_3$,label.angle=180}{t1}
\fmfv{label=$\alpha_2$,label.angle=0}{t3}
\end{fmfgraph*}\end{center}}}
\eeq
where $F_{\al\al'}^{}\big[\begin{smallmatrix} \al_3 & \al_2\\ \al_4 & \al_1\end{smallmatrix}\big]$ 
is the fusion matrix of the CFT. The two relevant pants decompositions of the four-punctured 
sphere are illustrated in Figure~\ref{fig:F-move}.

\begin{figure}[t]
\centering
$$
\text{\parbox{1.5in}{\epsfig{file=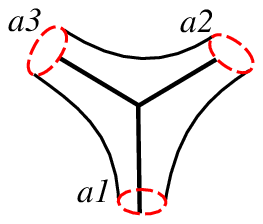,width=1.5in
\psfrag{a1}{$\alpha_1$}
\psfrag{a2}{$\alpha_2$}
\psfrag{a3}{$\alpha_3$}}}}
\ \longrightarrow
\ \text{\parbox{1.5in}{\epsfig{file=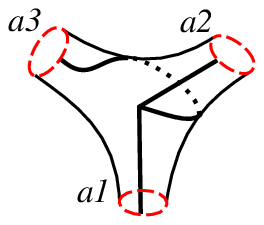,width=1.5in
\psfrag{a1}{$\alpha_1$}
\psfrag{a2}{$\alpha_2$}
\psfrag{a3}{$\alpha_3$}}}}
$$
\parbox{5in}{\caption{The braiding move exchanges orientation within one pair of pants.}
\label{fig:B-move}}
\end{figure}

Braiding, which exchanges two consecutive edges in the trivalent graph, 
acts by multiplication by a phase factor
\beq
\label{braiding-move}
\raisebox{0mm}[10mm][8mm]{\parbox{24mm}{\begin{center}
\begin{fmfgraph*}(15,15)
\fmfbottom{bl,br}
\fmftop{tl,tr}
\fmf{phantom}{bl,b1,b2,b3,br}
\fmf{phantom}{tl,t1,tt1,t2,tt3,t3,tr}
\fmffreeze
\fmf{plain,width=2}{t1,c2,t3}
\fmf{plain,width=2,tension=2}{c2,b2}
\fmfv{label=$\alpha_1$,label.angle=30}{b2}
\fmfv{label=$\alpha_2$,label.angle=180}{t1}
\fmfv{label=$\alpha_3$,label.angle=0}{t3}
\end{fmfgraph*}\end{center}}}
\equiv e^{i\pi(\Delta(\alpha_1)-\Delta(\alpha_2)-\Delta(\alpha_3))}
\raisebox{0mm}[10mm][8mm]{\parbox{24mm}{\begin{center}
\begin{fmfgraph*}(15,15)
\fmfbottom{bl,br}
\fmftop{tl,tr}
\fmf{phantom}{bl,b1,b2,b3,br}
\fmf{phantom}{tl,t1,tt1,t2,tt3,t3,tr}
\fmffreeze
\fmf{plain,width=2}{t1,c2,t3}
\fmf{plain,width=2,tension=2}{c2,b2}
\fmfv{label=$\alpha_1$,label.angle=30}{b2}
\fmfv{label=$\alpha_3$,label.angle=180}{t1}
\fmfv{label=$\alpha_2$,label.angle=0}{t3}
\end{fmfgraph*}\end{center}}}
\eeq
where $\Delta(\alpha)$ is the conformal dimension of the primary operator 
with quantum number (representation) $\alpha$. This is a bit clearer when 
drawn on the trinion, see Figure~\ref{fig:B-move}.

\begin{figure}[t]
\centering
$$
\text{\parbox{1.7in}{\epsfig{file=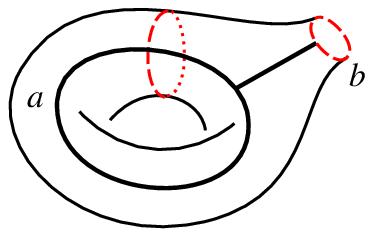,width=1.7in
\psfrag{a}{$\mu$}
\psfrag{b}{$\gamma$}}}}
\ \longrightarrow
\ \text{\parbox{1.7in}{\epsfig{file=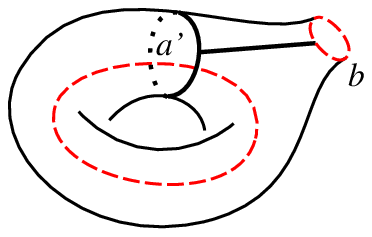,width=1.7in
\psfrag{a'}{$\alpha$}
\psfrag{b}{$\gamma$}}}}
$$
\parbox{5in}{\caption{The S-move exchanges two different pants-decompositions 
of the once-punctured torus and is represented by the matrix $S_{\mu \alpha}(\gamma)$. 
While the cycles carrying labels $\mu$ and $\alpha$ 
are different, the associated trivalent graphs look identical.}
\label{fig:S-move}}
\end{figure}

Finally, the $S$-move 
relates the following two trivalent graphs 
\beq
\label{S-move}
\hskip-5mm
\raisebox{0mm}[10mm][8mm]{\parbox{30mm}{
\begin{fmfgraph*}(30,30)
\fmfsurroundn{o}{8}
\fmf{phantom,tension=.5}{o1,v1}
\fmf{phantom,tension=.5}{o2,v2}
\fmf{phantom,tension=.5}{o3,v3}
\fmf{phantom,tension=.5}{o4,v4}
\fmf{phantom,tension=.5}{o5,v5}
\fmf{phantom,tension=.5}{o6,v6}
\fmf{phantom,tension=.5}{o7,v7}
\fmf{phantom,tension=.5}{o8,v8}
\fmfcyclen{plain,width=2,right=0.20}{v}{8} 
\fmffreeze
\fmf{plain,width=2}{o1,v1}
\fmf{phantom}{o4,o45,o5,o55,o6}
\fmffreeze
\fmf{phantom}{o1,c1,v1}
\fmfv{label=$\mu$,label.angle=0,label.dist=6}{v5}
\fmfv{label=$\gamma$,label.angle=-90,label.dist=6}{c1}
\end{fmfgraph*}}}
\ 
=\int d\alpha\,S_{\mu\alpha}(\gamma)
\hskip-5mm
\raisebox{0mm}[10mm][8mm]{\parbox{25mm}{
\begin{fmfgraph*}(30,30)
\fmfsurroundn{o}{8}
\fmf{phantom,tension=.5}{o1,v1}
\fmf{phantom,tension=.5}{o2,v2}
\fmf{phantom,tension=.5}{o3,v3}
\fmf{phantom,tension=.5}{o4,v4}
\fmf{phantom,tension=.5}{o5,v5}
\fmf{phantom,tension=.5}{o6,v6}
\fmf{phantom,tension=.5}{o7,v7}
\fmf{phantom,tension=.5}{o8,v8}
\fmfcyclen{plain,width=2,right=0.20}{v}{8} 
\fmffreeze
\fmf{plain,width=2}{o1,v1}
\fmf{phantom}{o4,o45,o5,o55,o6}
\fmffreeze
\fmf{phantom}{o1,c1,v1}
\fmfv{label=$\alpha$,label.angle=0,label.dist=6}{v5}
\fmfv{label=$\gamma$,label.angle=-90,label.dist=6}{c1}
\end{fmfgraph*}}}
\eeq
where $S_{\mu\alpha}(\gamma)$ is the modular matrix on the once-punctured torus. 
While the trivalent graphs on the left and right hand sides look identical, they actually 
represent different cycles on the once-punctured torus, see Figure~\ref{fig:S-move}.

\begin{figure}[t]
\centering
$$
\text{\parbox{1.2in}{\epsfig{file=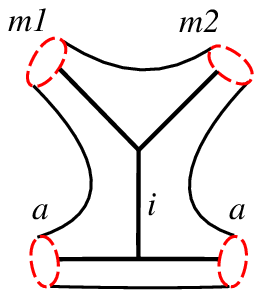,width=1.2in
\psfrag{m1}{$\mu_1$}
\psfrag{m2}{$\mu_2$}
\psfrag{i}{$\eta$}
\psfrag{a}{$\alpha$}
}}}
\ \to
\text{\parbox{1.2in}{\epsfig{file=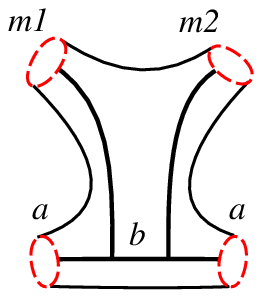,width=1.2in
\psfrag{m1}{$\mu_1$}
\psfrag{m2}{$\mu_2$}
\psfrag{b}{$\beta$}
\psfrag{a}{$\alpha$}
}}}
\ \to
\text{\parbox{1.2in}{\epsfig{file=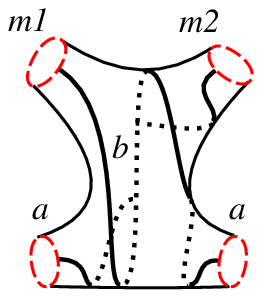,width=1.2in
\psfrag{m1}{$\mu_1$}
\psfrag{m2}{$\mu_2$}
\psfrag{b}{$\beta$}
\psfrag{a}{$\alpha$}
}}}
\ \to
\text{\parbox{1.2in}{\epsfig{file=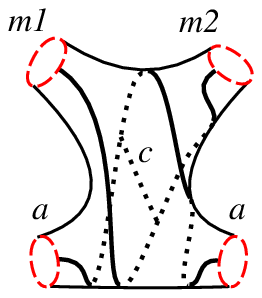,width=1.2in
\psfrag{m1}{$\mu_1$}
\psfrag{m2}{$\mu_2$}
\psfrag{c}{$\gamma$}
\psfrag{a}{$\alpha$}
}}}
$$
$$
\raisebox{0mm}[10mm][9mm]{\parbox{24mm}{\begin{center}
\begin{fmfgraph*}(20,15)
\fmfbottom{bl,br}
\fmftop{tl,tr}
\fmf{plain,width=2}{bl,b1,b2,b3,br}
\fmf{phantom}{tl,t1,t2,t3,tr}
\fmffreeze
\fmf{plain,width=2}{t1,c2,t3}
\fmf{plain,width=2}{c2,b2}
\fmfv{label=$\alpha$,label.angle=-90,label.dist=8}{bl}
\fmfv{label=$\alpha$,label.angle=-90,label.dist=8}{br}
\fmfv{label=$\mu_1$,label.angle=180}{t1}
\fmfv{label=$\mu_2$,label.angle=0}{t3}
\fmfv{label=$\eta$,label.angle=-45,label.dist=8}{c2}
\end{fmfgraph*}\end{center}}}
\to
\raisebox{0mm}[10mm][9mm]{\parbox{24mm}{\begin{center}
\begin{fmfgraph*}(20,15)
\fmfbottom{bl,br}
\fmftop{tl,tr}
\fmf{plain,width=2}{bl,b1,b2,b3,br}
\fmf{phantom}{tl,t1,t2,t3,tr}
\fmffreeze
\fmf{plain,width=2}{t1,b1}
\fmf{plain,width=2}{t3,b3}
\fmfv{label=$\alpha$,label.angle=-90,label.dist=8}{bl}
\fmfv{label=$\alpha$,label.angle=-90,label.dist=8}{br}
\fmfv{label=$\beta$,label.angle=-90,label.dist=8}{b2}
\fmfv{label=$\mu_1$,label.angle=180}{t1}
\fmfv{label=$\mu_2$,label.angle=0}{t3}
\end{fmfgraph*}\end{center}}}
\to
\raisebox{-6mm}[10mm][12mm]{\parbox{24mm}{\begin{center}
\begin{fmfgraph*}(20,30)
\fmfbottom{bl,br}
\fmftop{tl,tr}
\fmf{phantom}{bl,cl,tl}
\fmf{phantom}{br,cr,tr}
\fmffreeze
\fmf{plain,width=2}{cl,c1,c2,c3,cr}
\fmf{phantom}{tl,t1,t2,t3,tr}
\fmf{phantom}{bl,b1,b2,b3,br}
\fmffreeze
\fmf{plain,width=2}{t1,c1}
\fmf{plain,width=2}{b3,c3}
\fmfv{label=$\alpha$,label.angle=-90,label.dist=8}{cl}
\fmfv{label=$\alpha$,label.angle=-90,label.dist=8}{cr}
\fmfv{label=$\beta$,label.angle=-90,label.dist=8}{c2}
\fmfv{label=$\mu_1$,label.angle=180}{t1}
\fmfv{label=$\mu_2$,label.angle=30}{b3}
\end{fmfgraph*}\end{center}}}
\to
\raisebox{0mm}[10mm][9mm]{\parbox{24mm}{\begin{center}
\begin{fmfgraph*}(20,15)
\fmfbottom{bl,br}
\fmftop{tl,tr}
\fmf{plain,width=2}{bl,b1,b2,b3,br}
\fmf{phantom}{tl,t1,t2,t3,tr}
\fmffreeze
\fmf{plain,width=2}{t1,b1}
\fmf{plain,width=2}{t3,b3}
\fmfv{label=$\alpha$,label.angle=-90,label.dist=8}{bl}
\fmfv{label=$\alpha$,label.angle=-90,label.dist=8}{br}
\fmfv{label=$\beta$,label.angle=-90,label.dist=8}{b2}
\fmfv{label=$\mu_1$,label.angle=180}{t1}
\fmfv{label=$\mu_2$,label.angle=0}{t3}
\end{fmfgraph*}\end{center}}}
\to
\raisebox{0mm}[10mm][9mm]{\parbox{24mm}{\begin{center}
\begin{fmfgraph*}(20,15)
\fmfbottom{bl,br}
\fmftop{tl,tr}
\fmf{plain,width=2}{bl,b1,b2,b3,br}
\fmf{phantom}{tl,t1,t2,t3,tr}
\fmffreeze
\fmf{plain,width=2}{t1,c2,t3}
\fmf{plain,width=2}{c2,b2}
\fmfv{label=$\alpha$,label.angle=-90,label.dist=8}{bl}
\fmfv{label=$\alpha$,label.angle=-90,label.dist=8}{br}
\fmfv{label=$\mu_1$,label.angle=180}{t1}
\fmfv{label=$\mu_2$,label.angle=0}{t3}
\fmfv{label=$\gamma$,label.angle=-45,label.dist=8}{c2}
\end{fmfgraph*}\end{center}}}
$$
\parbox{5in}{\caption{The steps involved in calculating the Verlinde loop operator are 
a fusion, a double braiding and another fusion. They are represented in detail on the 
4-punctured sphere above and schematically below.}
\label{fig:wilson}}
\end{figure}

We are now ready to compute the action of ${\cal L}_\mu(p)$ 
on the conformal blocks ${\cal F}^{(\sigma)}_{\alpha,E}$ when 
$p$ is a path encircling one 
of the internal edges of the corresponding trivalent
graph $\Gamma_\sigma$. The pictorial representation capturing the 
monodromy is given in Figure~\ref{fig:wilson}. In the figure we have
allowed for arbitrary external states $\mu_1$ and $\mu_2$ and internal initial 
and final states $\eta$ and $\gamma$. For the purpose of computing the Verlinde 
loop operator, we will take $\eta$ and $\gamma$ to be the identity representation 
and therefore%
\footnote{Recall that the conventional orientation is such that the bottom left 
puncture of the four-point conformal block is mapped to infinity and that all states 
are oriented to the bottom and left, which is 
why it carries the label $\alpha$ and not $\alpha^*$.}
$\mu_2^*=\mu_1=\mu$. 
Then it is also natural to glue the upper two punctures into a 
torus, as we illustrate below in Figure~\ref{fig:torus}.

The first move in Figure~\ref{fig:wilson} is a fusion, expressing the t-channel conformal 
blocks in terms of the s-channel ones. The next steps, which are the hardest to visualize, 
are two braiding moves on the pair of pants labeled by representations $\beta$, $\alpha$ and $\mu_2$. 
Lastly another fusion move returns us to a t-channel graph similar to the original one.
The combined action of these moves act on the conformal blocks on 
the surface $C_{g,n+2}$ with the extra two punctures. It generates the monodromy
\beq
\cF^{(\hat\sigma)}_{\alpha\eta,\{\mu_1\mu_2\}}
=
\sum_{\beta}
F_{\eta\beta^*}
\big[\begin{smallmatrix}
\alpha^* & \mu_1\\
\alpha^* &\mu_2
\end{smallmatrix}\big]
e^{2\pi i(\Delta(\beta)-\Delta(\alpha)-\Delta(\mu_2))}
F_{\beta\gamma}
\left[\begin{smallmatrix}
\mu_1 & \mu_2\\
\alpha & \alpha
\end{smallmatrix}\right]
\cF^{(\hat\sigma')}_{\alpha\gamma,\{\mu_1\mu_2\}}\,.
\label{monodromy1}
\eeq
In order to derive the action of the loop operator ${\cal L}_\mu(p)$ we need to project down 
to the original conformal blocks on $C_{g,n}$, so we impose in the initial and final 
configurations in Figure~\ref{fig:wilson} that the 
intermediate channel carries the identity representation $\eta=\gamma=1$. 
We get 
the action of the Verlinde loop operator on the conformal blocks of the CFT on $C_{g,n}$%
\footnote{In defining the Verlinde loop operator, there is a normalization 
freedom by an arbitrary function of $\mu$. We include here the factor 
of $d_\mu=S_{1\mu}/S_{11}$, which is the quantum dimension of the 
representation labeled by $\mu$. 
This is the normalization convention used in \cite{Alday:2009fs}. 
When using the Verlinde loop operator to calculate Wilson loop operators in the 4d 
gauge theory, the difference is in the normalization of the trace. This 
convention gives for a trivial Wilson loop operator the dimension of 
the representation. See Section~\ref{sec:ZZ-loops}.
The convention in \cite{Drukker:2009id} did not include this factor, 
so that the VEV of the trivial Wilson loop was 1.\label{foot:normalization}}
\beq
[\cL_\mu(p)\cdot\cF^{(\sigma)}]_\alpha
=
d_\mu\sum_{\beta}
F_{1\beta^*}
\big[\begin{smallmatrix}
\alpha^* & \mu\\
\alpha^* &\mu^*
\end{smallmatrix}\big]
e^{2\pi i(\Delta(\beta)-\Delta(\alpha)-\Delta(\mu))}
F_{\beta1}
\left[\begin{smallmatrix}
\mu & \mu^*\\
\alpha & \alpha
\end{smallmatrix}\right]\cF^{(\sigma)}_\alpha\,.
\label{insertverlinde}
\eeq
This is to be contrasted with the effect of the topological loop operator ${\cal O}_\mu(p)$, 
which inserts expression (\ref{inserttopo})
\beq
{S_{\mu \alpha}\over S_{1\alpha}}\,.
\label{inserttopob}
\eeq
Is there any relation between the two expressions?

It turns out that the monodromy in (\ref{insertverlinde}) is identical to (\ref{inserttopob}).
This follows from the fact that the fusion, braiding and modular matrices of a CFT are not
independent, rather they satisfy the so-called Moore-Seiberg groupoid relations 
\cite{Moore:1988qv}. These relations
can be obtained by manipulating trivalent graphs for the cases $g=0$ with $n=4,5$ as
well as $g=1$ with $n=1,2$.

\begin{figure}[t]
\centering
$$
\text{\parbox{1in}{\epsfig{file=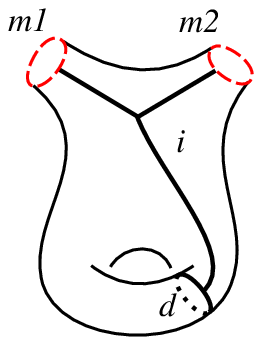,width=1in
\psfrag{m1}{$\mu_1$}
\psfrag{m2}{$\mu_2$}
\psfrag{i}{$\eta$}
\psfrag{d}{$\delta$}
}}}
\ \to
\text{\parbox{1in}{\epsfig{file=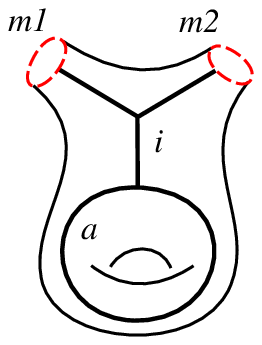,width=1in
\psfrag{m1}{$\mu_1$}
\psfrag{m2}{$\mu_2$}
\psfrag{i}{$\eta$}
\psfrag{a}{$\alpha$}
}}}
\ \to\cdots\to
\text{\parbox{1in}{\epsfig{file=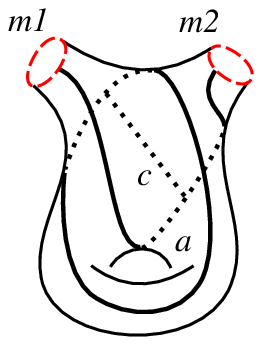,width=1in
\psfrag{m1}{$\mu_1$}
\psfrag{m2}{$\mu_2$}
\psfrag{c}{$\gamma$}
\psfrag{a}{$\alpha$}
}}}
$$
\parbox{5in}{\caption{When connecting the two bottom punctures in Figure~\ref{fig:wilson}, 
we get a twice punctured torus. In order to derive the relation \eqn{relation} we act first by 
one further S-move.}
\label{fig:relation-1}}
\end{figure}

\begin{figure}[t]
\centering
$$
\text{\parbox{1in}{\epsfig{file=relation1.eps,width=1in
\psfrag{m1}{$\mu_1$}
\psfrag{m2}{$\mu_2$}
\psfrag{i}{$\eta$}
\psfrag{d}{$\delta$}
}}}
\ \to
\text{\parbox{1in}{\epsfig{file=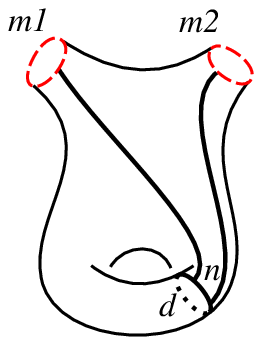,width=1in
\psfrag{m1}{$\mu_1$}
\psfrag{m2}{$\mu_2$}
\psfrag{d}{$\delta$}
\psfrag{n}{$\nu$}
}}}
\ \to
\text{\parbox{1in}{\epsfig{file=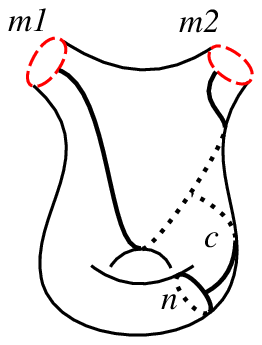,width=1in
\psfrag{m1}{$\mu_1$}
\psfrag{m2}{$\mu_2$}
\psfrag{c}{$\gamma$}
\psfrag{n}{$\nu$}
}}}
\ \to
\text{\parbox{1in}{\epsfig{file=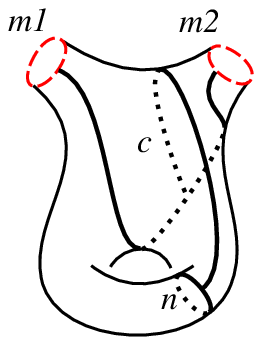,width=1in
\psfrag{m1}{$\mu_1$}
\psfrag{m2}{$\mu_2$}
\psfrag{c}{$\gamma$}
\psfrag{n}{$\nu$}
}}}
\ \to
\text{\parbox{1in}{\epsfig{file=relation6.eps,width=1in
\psfrag{m1}{$\mu_1$}
\psfrag{m2}{$\mu_2$}
\psfrag{c}{$\gamma$}
\psfrag{a}{$\alpha$}
}}}
$$
\parbox{5in}{\caption{The same result as in Figure~\ref{fig:relation-1} 
is achieved by doing two fusion steps, a braiding and an S-move.}
\label{fig:relation-2}}
\end{figure}

The relation that is relevant for us is one that arises for the 
twice-puntured torus ($g=1, n=2$) (see \cite{Behrend:1999bn} for a review). 
It is illustrated in Figures~\ref{fig:relation-1} and~\ref{fig:relation-2}. 
We use the twice-punctured torus, where cutting the torus would give the 
four punctured sphere used in the calculation of the Verlinde loop operator 
in Figure~\ref{fig:wilson}.
As a first step we take an S-move $S_{\delta\alpha}(\eta)$ and then continue 
with the steps of the Verlinde loop operator, which are a fusion, two braidings 
and a further fusion. 

The same result is achieved by a different sequence of steps in 
Figure~\ref{fig:relation-2}. There are first two fusions, then a braiding, and finally an S-move. 
These steps are represented by the monodromy
\beq
\sum_{\nu}
F_{\eta\nu}
\left[\begin{smallmatrix}
\delta & \mu_1\\
\delta &\mu_2
\end{smallmatrix}\right]
F_{\delta^*\gamma}
\left[\begin{smallmatrix}
\mu_2 & \mu_1 \\
\nu^* & \nu^*
\end{smallmatrix}\right]
e^{\pi i(\Delta(\mu_1)-\Delta(\mu_2)-\Delta(\gamma))}
S_{\nu\alpha}(\gamma)
\,.
\label{relation}
\eeq
This is therefore equal to the expression in \eqn{monodromy1} multiplied by 
$S_{\delta\alpha}(\eta)$.

Specializing to our case, when $\eta=\gamma=1$ we get the relation
\beq
F_{1\beta^*}
\big[\begin{smallmatrix}
\alpha^* & \mu\\
\alpha^* &\mu^*
\end{smallmatrix}\big]
e^{2\pi i(\Delta(\beta)-\Delta(\alpha)-\Delta(\mu))}
F_{\beta1}
\left[\begin{smallmatrix}
\mu & \mu^*\\
\alpha & \alpha
\end{smallmatrix}\right]
S_{\delta \alpha}
=
\sum_{\nu}
F_{1\nu}
\left[\begin{smallmatrix}
\delta & \mu\\
\delta &\mu^*
\end{smallmatrix}\right]
F_{\delta^*1}
\left[\begin{smallmatrix}
\mu^* & \mu \\
\nu^* & \nu^*
\end{smallmatrix}\right]
S_{\nu\alpha}\,.
\label{two-torus}
\eeq
where $S_{\delta\alpha}\equiv S_{\delta\alpha}(1)$. We now further restrict to the case 
when $\delta=1$, which imposes $\nu=\mu$ and eliminates the sum on the 
right hand side of (\ref{two-torus}). 
Using the expression for $d_\mu$, the {\it quantum dimension} of the representation 
labeled by $\mu$
\beq
F_{11}
\big[\begin{smallmatrix}
\mu^* & \mu\\
\mu^* & \mu^*
\end{smallmatrix}\big]
={1\over d_\mu}={S_{11}\over S_{1\mu}}\,,
\eeq
we finally find
\beq
d_\mu
\sum_{\beta}
F_{1\beta^*}
\big[\begin{smallmatrix}
\alpha^* & \mu\\
\alpha^* &\mu^*
\end{smallmatrix}\big]
e^{2\pi i(\Delta(\beta)-\Delta(\alpha)-\Delta(\mu))}
F_{\beta1}
\left[\begin{smallmatrix}
\mu & \mu^*\\
\alpha & \alpha
\end{smallmatrix}\right]
=
\frac{S_{\mu\alpha}}{S_{1\alpha}}\,,
\label{equality}
\eeq
which is precisely the expression for the topological defect.

Since, topological defect operators and Verlinde loop operators transform 
the same way under elements of the Moore-Seiberg groupoid ${\cal G}$, 
their equivalence extends to an arbitrary choice of pants decomposition.

We conclude that topological defect operators and Verlinde loop operators 
are isomorphic:
\beq
{\cal O}_\mu(p)={\cal L}_\mu(p)\,.
\label{proof}
\eeq
This result, apart from identifying two seemingly different classes of operators in 
2d CFTs, provides a very effective tool for computing the correlation functions of 
loop operators. The fusion and braiding matrices of a CFT are given by complicated 
expressions involving quantum deformations of $6j$ symbols and are unknown for most CFTs. 
However, our identification reduces the correlators to the computation of the modular 
matrices of the CFT, which can be much more easily derived (see Section~\ref{sec:topliouville}).

Let us make one more observation. If we connect the two upper 
punctures of the four-punctured sphere (which is quite natural, since they carry 
conjugate representations) the starting configuration has the state $\mu$ running around the 
torus and the state $\alpha$ along a line between the two punctures, with nothing 
(i.e. the identity state) connecting them. After the monodromy we get a similar 
configuration, only that now, if viewed in an embedding 3d space, the torus 
wraps the line. This is illustrated in Figure~\ref{fig:torus} and is closely related to the realization 
of topological defects using topological field theories 
(see e.g. \cite{Witten-CS,Felder:1999cv, Fuchs:2002cm}). If the CFT has a 3d topological 
field theory realization, then the topological defect operators are realized as closed 
curves in the bulk that are linked to the curves associated to the local operators,\footnote{This observation was recently made in the context of Liouville theory in \cite{Wu:2009tq}.} 
which are represented by curves ending on the boundary.

\begin{figure}[t]
\centering
$$
\begin{tabular}{ccp{.3in}cc}
\text{\parbox{1.2in}{\epsfig{file=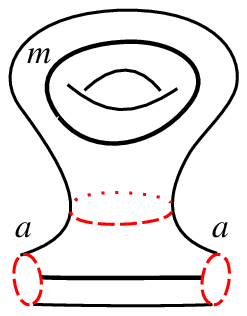,width=1.2in
\psfrag{m}{$\mu$}
\psfrag{a}{$\alpha$}
}}}
&
\text{\parbox{1.2in}{\epsfig{file=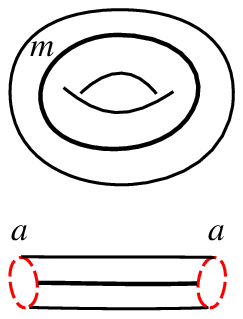,width=1.2in
\psfrag{m}{$\mu$}
\psfrag{a}{$\alpha$}
}}}
&&
\text{\parbox{1.2in}{\epsfig{file=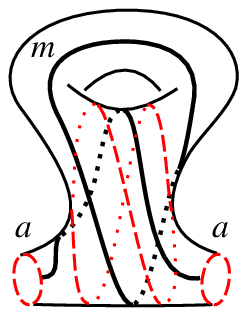,width=1.2in
\psfrag{m}{$\mu$}
\psfrag{a}{$\alpha$}
}}}
&
\text{\parbox{1.2in}{\epsfig{file=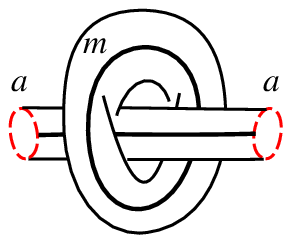,width=1.2in
\psfrag{m}{$\mu$}
\psfrag{a}{$\alpha$}
}}}
\\[3mm]
$(a)$&$(b)$&&$(c)$&$(d)$
\end{tabular}
$$
\parbox{5in}{\caption{With the identity state in the intermediate channel, the starting 
point $(a)$ is like the addition of a disconnected torus carrying the state $\mu$ $(b)$. 
Doing the same thing for the final state, where the identity is along the channel 
marked by the dashed red line $(c)$ gives back a disconnected torus, but now 
winding around the original line.}
\label{fig:torus}}
\end{figure}

\begin{figure}[t]
\centering
\epsfig{file=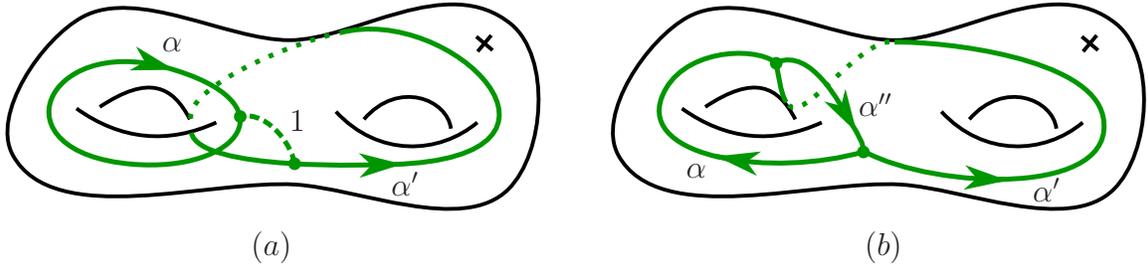,width=6in
\psfrag{c}{$\alpha$}
\psfrag{i}{$1$}
\psfrag{c'}{$\alpha'$}
\psfrag{c''}{$\alpha''$}}
$(a)\hskip3in(b)$
\parbox{5in}{\caption{A pair of intersecting topological defects $(a)$ can be 
resolved (after including the extra trivial line carrying the identity state) to 
a defect web with no intersections, but with two junctions $(b)$.}
\label{fig:skein}}
\end{figure}

This approach makes quite clear that topological line operators and webs will 
satisfy properties analogous to the properties of the conformal block trivalent graphs. 
For example, a web containing four strands with labels $\alpha_i$ connected through 
two trivalent vertices by a strand of label $\alpha$ in an ``s-channel'' configuration 
or similar web where the four strands are connected through a strand of label 
$\tilde \alpha$ in a ``t-channel'' configuration will be related through a fusion 
matrix coefficient exactly as in eqn. (\ref{genfus})! This is illustrated in Figure~\ref{fig:skein}.

Also, we learn how to ``resolve'' the intersection between the webs ${\cal O}$ and ${\cal O}'$ 
in ${\cal O} \circ {\cal O}'$. If we focus on the region near an intersection, and add a 
line carrying the identity representation and joining the two crossing strands, the 
intersection looks like a ``u-channel'' four point conformal block. Through a 
fusion and a braiding operation, this can be resolved to a web with no self-intersections. 
Hence we have an algebra
\beq
{\cal O}[\alpha] \circ {\cal O}' [\alpha'] = \int d\alpha'' C(\alpha, \alpha', \alpha'') {\cal O}''[\alpha'']\,.
\eeq
Here $\alpha$,$\alpha'$,$\alpha''$ denote the labels of representations on the webs 
${\cal O}$,${\cal O}'$,${\cal O}''$ respectively.

The skein relations for loop operators in Liouville CFT discussed in 
\cite{Drukker:2009id} are an example of this operation. In that case, when the 
lines carry the first degenerate representation, they can be resolved into a 
linear combination of a ``t-channel'' and ``s-channel'' graph with a trivial 
intermediate state, {\it i.e.}, into pairs of non-intersecting curves.

\section{Loop Operators in Liouville Theory and Toda}
\label{sec:topliouville}
\setcounter{subsubsection}{0}

In this section we study topological defect operators in an interesting class of non-rational 
CFTs. We start with an analysis of topological defects in Liouville theory and then explicitly 
construct topological defects in Toda CFTs. We find that these operators are labeled by 
representations (degenerate and non-degenerate) of the Virasoro algebra.

In order to compute the topological defects in Toda CFTs we first calculate the 
characters for arbitrary representations in Section~\ref{sec:Toda-reps} and present 
the explicit formula for the operators in Section~\ref{sec:charactersss}. In 
Section~\ref{sec:topjanus} we construct an interesting class of generalized topological 
line operators, which when inserted in a CFT correlator, effectively change the 
complex structure moduli of the Riemann surface $C_{g,n}$ where a CFT correlator is defined 
in the holomorphic (or antiholomorphic) sector. Insertion of these loop operators 
yield CFT correlators where the holomorphic and antiholomorphic sectors are 
based on different Riemann surfaces.

\subsection{Topological Defect Operators in Liouville CFT}

Topological defect operators in Liouville can be obtained via the unfolding
trick from the boundary states of $\text{Liouville}\otimes \overline{\text{Liouville}}$.
Boundary states in Liouville theory 
\cite{Fateev:2000ik,Teschner:2000md,Zamolodchikov:2001ah} 
are labeled by the representations of the
Virasoro algebra, akin to the Cardy classification of boundary states in RCFTs
in terms of representations of the chiral algebra ${\cal A}$.

A representation of the Virasoro algebra is characterized by the Liouville 
momentum $\alpha$ of the primary field, which has dimension 
$\Delta(\alpha)=\alpha(Q-\alpha)$, where $Q$ determines the CFT central 
charge through $c=1+6Q^2$. There are two families of boundary 
states in Liouville theory labeled by different representations of the Virasoro algebra:

\begin{itemize}
\item ZZ boundary states \cite{Zamolodchikov:2001ah}. Labeled by a pair of integers 
$(r,s)\in \bZ_{+}\times \bZ_{+}$ describing a {\it degenerate representation},%
\footnote{An $(r,s)$ degenerate representation contains a Virasoro descendant 
at level $rs$ that is a null state.} 
with momentum
\beq
2\alpha_{r,s}=Q-\frac{r}{ b} -s b\,,
\label{momdege}
\eeq
where $Q=b+1/b$. The corresponding character is given by
\beq
\chi_{r,s}(\tau)={q^{-(r/b+sb)^2/4}-q^{-(r/b-sb)^2/4}\over \eta(\tau)}\,.
\label{characterden}
\eeq

\item FZZT boundary states \cite{Fateev:2000ik,Teschner:2000md}. Labeled 
by a real parameter $m\in i\bR$ 
describing a non-degenerate representation with momentum
\beq
\mu ={Q\over 2}+m\,.
\label{liouville-nondeg}
\eeq
The corresponding character is given by
\beq
\chi_{\mu}(\tau)={q^{-m^2}\over \eta(\tau)}\,.
\label{characternonden}
\eeq

\end{itemize}

Recently, an investigation of topological defect operators in Liouville theory has been
carried out in \cite{Sarkissian:2009aa} by unfolding the permutation boundary states
in $\text{Liouville}\otimes \overline{\text{Liouville}}$. These boundary states, which where obtained 
by extending the boundary conformal bootstrap approach in 
\cite{Fateev:2000ik, Teschner:2000md,Zamolodchikov:2001ah}, yield defect operators 
also classified by degenerate and non-degenerate representations of the Virasoro algebra:
\begin{itemize}
\item ZZ topological defect:
\beq
{\cal O}_{r,s}(p)
=\int_{{Q\over 2}+i\bR} d\alpha\, D_{r,s}(\alpha) 
\sum_{\{k\},\{l\}} |\alpha,k;\alpha^*,l \rangle\otimes \langle \alpha,k;\alpha^*,l|\,,
\label{ZZ}
\eeq
where 
\beq
D_{r,s}(\alpha)= \frac{\sin(2\pi ra/b)\sin(2\pi s ab)}{ \sin(2\pi a/b)\sin(2\pi ab)}\,,
\label{ZZa}
\eeq
and $a=\alpha-Q/2$.

\item FZZT topological defect:
\beq
{\cal O}_{\mu}(p)=\int_{{Q\over 2}+i\bR} d\alpha\, D_\mu(\alpha) 
\sum_{\{k\},\{l\}} |\alpha,k;\alpha^*,l \rangle\otimes \langle \alpha,k;\alpha^*,l|\,,
\label{FZZT}
\eeq
where \beq
D_{\mu}(\alpha)=-\frac{\cos(4\pi am) }{ 2 \sin(2\pi a/b)\sin(2\pi ab)}\,.
\label{FZZTa}
\eeq
and $m=\mu-Q/2$.
\end{itemize}

We now show that despite that Liouville theory is not a RCFT, the expression 
for the topological defects can be written in terms of the modular matrices of 
the corresponding characters as in eqn. (\ref{modular}). 
Performing the 
modular transformation of the degenerate characters (\ref{characterden}) 
and non-generate characters (\ref{characternonden}) the following modular 
matrices can be derived \cite{Zamolodchikov:2001ah} 
\beq
S_{(r,s) \alpha}=-2\sqrt{2}i\sin(2\pi a r/b)\sin(2\pi a sb)\,
\eeq
and
\beq
S_{\mu \alpha}=\sqrt{2}i \cos(4\pi m a)\,.
\eeq
Therefore, by comparing with (\ref{ZZa})(\ref{FZZTa}) we find that
\beq
D_{r,s}(\alpha)={S_{(r,s) \alpha}\over S_{1a}}\qquad 
D_{\mu}(\alpha)={S_{\mu \alpha}\over S_{1\alpha}}\,.
\eeq

This computation highlights the simplicity of topological defect operators 
(or Verlinde loop operators) even for non-rational CFTs.

\subsection{States in Toda CFTs and Characters}
\label{sec:Toda-reps}
Toda CFTs are based on a Lie algebra $\gfrak$, with the choice 
$\gfrak=A_1$ corresponding to Liouville theory. ADE Toda CFTs 
are conjectured to capture observables in four dimensional gauge 
theories based on ADE gauge groups. Here we construct 
topological defect operators in these theories. But first we must construct the 
characters of arbitrary representations in Toda CFTs.

Toda CFTs have a W-algebra symmetry. This is generated by $K-1$ currents $W^{(l_i+1)}$
with spin $l_i+1$, 
where the set 
$\{l_i\}_{i=1,\ldots,K-1}$ is the set of exponents\footnote{The set of 
exponents $\{l_i\}$ of $\gfrak$ is related to the set of the order $\{\nu_i\}$ 
of the Casimirs of $\gfrak$ by $l_i=\nu_i-1$.} of $\gfrak$ and $K-1=\hbox{rank}(\gfrak)$.
The generator $W^{(2)}=T$ denotes the usual energy-momentum tensor 
of a CFT and generates the Virasoro subalgebra ${\it Vir}\subset W$.

Vertex operators of primary operators in Toda CFTs are constructed out of 
$K-1$ scalars, and are given by
\beq
e^{\langle \mu, \phi\rangle}\,,
\label{vertex}
\eeq
where $\langle~, ~\rangle$ is the bilinear form on $\hfrak^*$, the dual 
to the Cartan subalgebra $\hfrak$ of $\gfrak$. 
The conformal dimension of the primary operator (\ref{vertex}) is given by
\beq
\Delta(\mu)=\langle Q,\mu\rangle-{1\over 2}\langle\mu,\mu\rangle\,.
\label{delta-toda}
\eeq
$Q$ is the background charge for the $K-1$ scalars 
\beq 
Q=b\rho+{\rho^* \over b}\,,
\eeq
where $\rho$ is the Weyl vector\footnote{The Weyl vector is the sum over 
positive roots of $\gfrak$; that is $\rho=1/2\sum_{e>0} e$, while the dual Weyl vector 
$\rho^*=1/2\sum_{e^*>0} e^*$ is the sum over positive coroots.} of $\gfrak$ and $\rho^*$ is the dual 
Weyl vector. The background charge determines the central charge of the CFT through
\beq
c=K-1+12 \langle Q ,Q\rangle\,.
\eeq

There is a variety of possible representations in Toda CFTs. 
The spectrum of delta function normalizable states in Toda CFT is given 
by the non-degenerate representations. These are generated by primary states with%
\footnote{For $K=2$ this agrees with \eqn{liouville-nondeg}, only that in Liouville 
theory different conventions are used and in particular 
$Q_\text{Liouville}=2Q_\text{Toda}$.}
\beq
\mu=Q+m\,,
\label{nondegtoda}
\eeq
where $m$ is an arbitrary $K-1$ dimensional vector with imaginary entries.
Notice that momenta related by the Weyl group ${\cal W}$ action $m \to w(m)$, where $w\in {\cal W}$, 
correspond to the same representation. 
These representations consist of a full Verma module built 
out of the W-algebra descendants of the highest weight vector. 

To calculate the character we note that the states created by the W-algebra 
generators are just like those created by the free action of $K-1$ copies of the Virasoro 
generators. Therefore using \eqn{delta-toda} we get the characters for these 
representations
\beq
\chi_{\mu}(\tau)=\Tr\left(q^{L_0-c/24}\right)
=\left(\frac{q^{1/24}}{\eta(\tau)}\right)^{K-1}q^{\Delta(\mu)-c/24}
={q^{-{1\over 2}\langle m,m\rangle}\over \eta(\tau)^{K-1}}\,.
\label{nongenerate}
\eeq

In addition to these non-degenerate representations, other interesting representations 
in Toda CFTs can be constructed which contain null descendant states, known as 
semi-degenerate representations. 
A null vector at level $rs$  in the Verma module  is visible in the Kac determinant (see eqn. (6.72) in \cite{Bouwknegt:1992wg}), together with a whole Verma module of 
descendants, 
if there are positive integers $r$ and $s$ such that 
the inner product of the Toda momentum with a root $e\in \gfrak$ satisfies
\beq
-\langle m,e \rangle = {1\over 2} \langle e,e\rangle r b + s/b\,.
\eeq
Without loss of generality, we can classify 
semi-degenerate representations by a collection of simple roots $\ci$,\footnote{It will 
become apparent that two collections of simple roots related by the action of the Weyl 
group ${\cal W}$ give rise to the same representation.} 
and a set of conditions 
\beq 
-\langle m,e_i \rangle = {1\over 2} \langle e_i,e_i\rangle r_i b + s_i /b \qquad i \in \ci\,.
\label{semi-deg-weights}
\eeq
The components of the momentum in directions orthogonal 
to the simple roots in $\ci$ are free to assume arbitrary imaginary 
value $\tilde m$ (and as before, they are identified if related by a Weyl reflection). 
The simplest semi-degenerate representations have a single null vector at level $1$,
i.e. a momentum $\mu$ orthogonal to a single simple root. In other words, 
the highest weight vector is annihilated by some linear combination of the level $1$ 
W-algebra generators $\sum c_i W^{(l_i+1)}_{-1}$. The character of this 
representation takes the form 
\beq 
\chi^{(1)}_{\mu_{(1,1)}}(\tau)
={q^{-{1\over 2}\langle m,m\rangle}(1-q) \over \eta(\tau)^{K-1}} 
= {q^{-{1\over 2}\langle \tilde m,\tilde m\rangle}
(q^{-{1\over 4} (b+1/b)^2} - q^{-{1\over 4} (b-1/b)^2} ) \over \eta(\tau)^{K-1}}\,.
\eeq 
Similarly, the character of a representation with a single null vector at level $rs$ 
takes the form 
\beq 
\chi^{(1)}_{\mu_{(r,s)}}(\tau)
={q^{-{1\over 2}\langle m,m\rangle}(1-q^{rs}) \over \eta(\tau)^{K-1}} 
= {q^{-{1\over 2}\langle \tilde m,\tilde m\rangle}
(q^{-{1\over 4} (r b+s/b)^2} - q^{-{1\over 4} (r b-s/b)^2} ) \over \eta(\tau)^{K-1}}\,.
\eeq
In particular, for $K=2$ we recover the character of the degenerate representations of Liouville 
theory \eqn{characterden} (where $\tilde m=0$).

An extreme case are the fully degenerate representations, where $\cal I$ consists of {\em all} the 
simple roots in the Lie algebra $\gfrak$. The momentum of such a state
is constructed out of a weight vector and a coweight vector of $\gfrak$ multiplied 
by $b$ and $1/b$ respectively.\footnote{We recall 
that the fundamental weights of $\gfrak$ are dual to the simple coroots, while the 
fundamental coweights of $\gfrak$ are dual to the simple roots.}
Therefore, completely degenerate representations are labeled by two highest 
weight states $\lambda_1$ and $\lambda_2$, 
corresponding to a representation $R_1$ of $\gfrak$ and a representation\footnote{We recall
that a coweight of $\gfrak$ corresponds to a weight of $\gfrak^*$.} $\LR_2$ of $\Lgfrak$, 
the GNO or Langlands dual Lie algebra 
of $\gfrak$
\beq
m_{(R_1,\LR_2)}=-b(\rho+\lambda_1)-{1\over b}(\rho^*+\lambda_2)\,.
\label{completely}
\eeq
These representations have $K-1$ independent null vectors, which occur 
at Virasoro descendent level
\beq
r_is_i=\langle\rho+ \lambda_1,e_i^*\rangle
\langle e_i, \rho^*+\lambda_2\rangle,
\qquad 
i=1,\ldots,K-1\,.
\eeq

The simplest of the fully degenerate representations is the identity representation, 
for which $r_i=s_i=1$, so all the level $1$ vectors are null. 
Regularity of the operators 
$W^{(l_i+1)}$ at the origin of the complex plane imply the 
existence of the following null states%
\footnote{Out of these, $K-1$ are independent 
and generated by $\{W_{-1}^{(2)},W_{-1}^{(l_2+1)},\ldots, W_{-1}^{(l_{K-1}+1)}\}$.}
\beq
W_{-n}^{(l_i+1)}|0\rangle=0\,,\qquad 0< n\leq l_i\,.
\label{null}
\eeq
The irreducible Verma module associated to the vacuum is constructed 
by modding out by the descendants generated by the action of $W_{-n}^{(l_i+1)}$ 
for $0< n\leq l_i$. Therefore, the character is given by
\beq
\chi_1(\tau)={q^{-c/24}}\prod_{i=1}^{K-1}\prod^\infty_{n_i= l_i+1}{1\over 1-q^{n_i}}=
{q^{-{1\over 2}\langle Q, Q\rangle} \over \eta(\tau)^{K-1}}
\prod_{i=1}^{K-1} \prod_{n=1}^{l_i}(1-q^n)\,.
\label{identitychar}
\eeq
Using the Weyl denominator formula one can write this as follows
\beq
\chi_1(\tau)=\frac{{q^{-{1\over 2}\langle Q, Q\rangle}}}{\eta(\tau)^{K-1}}
\sum_{w\in {\cal W}} \epsilon(w) q^{-\left<w(\rho)-\rho, \rho^*\right>}\,.
\label{characsum}
\eeq
Here ${\cal W}$ is the Weyl group of $\gfrak$ and $w(\cdot)$ denotes the action 
of the element $w$ on a vector.

The extension of this formula to other completely degenerate representations 
\eqn{completely} 
is
\beq
\chi_{(R_1,\LR_2)}(\tau)={q^{\Delta(\mu)-(c-K+1)/24}\over \eta(\tau)^{K-1}}
\sum_{w\in {\cal W}} \epsilon(w) 
q^{-\left<w(\rho+\lambda_{1})-\rho-\lambda_{1}, \rho^*+\lambda_{2}\right>}\,.
\label{characsuma}
\eeq

It is even more subtle to compute the character for a generic semi-degenarate 
representation \eqn{semi-deg-weights}
in Toda theory as the Verma modules of different null vectors may intersect. 
We derive the characters by generalizing the examples above. 

For  a semi-degenerate representation, a subset $\cal I$ of the simple roots in $\gfrak$ 
has a distinguished role related to the null vectors. 
Consider the subgroup ${\cal W}_\ci$ of the Weyl group ${\cal W}$ of $\gfrak$ generated by 
reflections by the simple roots in $\ci$. 
Notice that ${\cal W}_\ci$ is the Weyl group of the subsystem of roots 
$\Delta_\ci$ which are linear combinations of the simple roots in $\ci$. 
We can define a restricted Weyl vector $\rho_\ci$ as half the sum of the 
positive roots in $\Delta_\ci$ and write 
$m = \tilde m - (\rho_\ci+\lambda_1)b - (\rho^*_\ci+\lambda_2)/b$, where $\vev{\tilde m, \rho_\ci}=\vev{\tilde m, \rho^*_\ci}=0$.
The vector $\lambda_1$ can be thought as a highest weight labeling an irreducible 
representation $R_1$ of the Lie algebra $\gfrak_\ci$ built from $\Delta_\ci$ and 
$\lambda_2$ as a highest weight labeling an irreducible 
representation $\LR_2$ of $\Lgfrak_\ci$, the GNO or Langlands dual Lie algebra $\gfrak_\ci$. 

Then the semi-degenerate character can be written as
\bal
\chi^\ci_{\tilde m,R_1,\LR_2}(\tau)
&=\frac{{q^{-{1\over 2}\langle \tilde m,\tilde m \rangle}}}{\eta(\tau)^{K-1}}
\sum_{w\in{\cal W}_\ci} \epsilon(w) 
q^{-{1\over 2} \langle w(\rho_\ci+\lambda_1) b + (\rho_\ci^*+\lambda_2)/b, 
w(\rho_\ci+\lambda_1) b + (\rho_\ci^*+\lambda_2)/b \rangle}
\\&
=\frac{{q^{-{1\over 2}\langle m,m \rangle}}}
{\eta(\tau)^{K-1}}\sum_{w\in{\cal W}_\ci} \epsilon(w) 
q^{\langle \rho_\ci+\lambda_1,\rho_\ci^*+\lambda_2\rangle
-\langle w(\rho_\ci+\lambda_1),\rho_\ci^*+\lambda_2\rangle}\,.
\label{characsemisum}
\eal
These formulae have a rather intuitive meaning. 
The term where $w$ is the identity represents the full 
Verma module. The terms where $w$ is a reflection by a simple root have 
$\epsilon(w) = -1$ and subtract off the Verma modules of the null vectors 
whose levels can be read from (\ref{semi-deg-weights}). 
The remaining terms deal with the intersections between the Verma modules. 
The formula is essentially fixed by invariance under ${\cal W}_\ci$.

When $\lambda_1 = \lambda_2=0$ we can use the Weyl denominator formula to rewrite 
the sum as 
\beq
\chi^\ci_{\tilde m}(\tau)=\frac{{q^{-{1\over 2}\langle m,m \rangle}}}
{\eta(\tau)^{K-1}} \prod_{\beta \in \Delta^+_\ci} (1- q^{\langle \beta,\rho^*_\ci\rangle})\,.
\label{characsemiprod}
\eeq
Since $\langle \beta,\rho_\ci\rangle$ simply counts the number of simple roots 
in $\beta$, we see the expected level $1$ singular vectors, and some more.

\subsection{Topological Defect Operators in Toda CFTs}
\label{sec:charactersss}
We now construct the topological defect operators corresponding to the various 
representations in Toda CFTs.
In order to determine the topological defect operator for a representation, 
we must first calculate the modular matrix of the associated character. An 
elementary computation yields\footnote{In deriving this formula we used that 
$\langle w(a),w(a)\rangle =\langle a,a\rangle$.}
\beq
\exp\left({i\pi\langle m,m\rangle\over \tau}\right)
=(i\tau)^{r/2} \sqrt{\hbox{det}(C)} \int d\alpha \, q^{-{1\over 2} \langle a,a\rangle}
{1\over |{\cal W}|}\sum_{w\in {\cal W}}\exp\left(2\pi i \langle w(m),a\rangle\right)\,,
\eeq
where $m=\mu-Q$ and $a=\alpha-Q$.

Using that 
$\eta(-1/\tau)=\sqrt{-i\tau} \eta(\tau)$, we arrive at the modular matrix for the 
non-degenerate representation labeled by $\mu$
\beq
S_{\mu \alpha}=i^{K-1} \sqrt{\hbox{det}(C)}{1\over |{\cal W}|}
\sum_{w\in {\cal W}} \exp\left(2 \pi i \langle w(m), a\rangle \right),
\label{Sma}
\eeq
where $C$ is the Cartan matrix of $\gfrak$.

In order to write down the formula for the topological defect (\ref{inserttopob}) 
we also need $S_{1\mu}$, obtained from the modular transformation of the 
character (\ref{characsum}) for the identity (vacuum) representation.
Each term in the sum in (\ref{characsum}) corresponds to the non-degenerate character 
(\ref{nongenerate}) evaluated at momentum
\beq
m=bw(\rho)+{1\over b} \rho^* \,.
\eeq
Therefore, 
\beq
S_{1\alpha}=i^{K-1} \sqrt{\hbox{det}(C)}{1\over |{\cal W}|}
\sum_{w\in {\cal W}}\epsilon(w)
\exp\bigg( 2\pi i b\langle w(\rho),a\rangle\bigg)
\sum_{w'\in {\cal W}}\epsilon(w')
\exp\left({2\pi i \over b}\langle w'(\rho^*),a\rangle\right)\,,
\eeq
which by Weyl's denominator formula can be written as
\beq
S_{1\alpha}=i^{K-1} \sqrt{\hbox{det}(C)}{1\over |{\cal W}|} 
\prod_{e>0}4\sin\bigg(\pi b \langle a,e\rangle\bigg)
\sin\bigg(-\frac{\pi}{b} \langle a,e^*\rangle\bigg)\,,
\eeq
where $e>0$ denote the positive roots of $\gfrak$, while $e^*>0$ are the positive coroots of $\gfrak$.

For the completely degenerate representations the character is given in equation 
\eqn{characsuma}. Each term in the sum corresponds to the character 
(\ref{nongenerate}) evaluated at momentum
\beq
m=bw(\rho+\lambda_1)+{1\over b} (\rho^*+\lambda_2)\,.
\eeq
Therefore, 
\bal
S_{(R_1,\LR_2)\alpha}=i^{K-1} \sqrt{\hbox{det}(C)}{1\over |{\cal W}|}
&\sum_{w\in {\cal W}}\epsilon(w)
\exp\left(2\pi i b\left< w(\rho+\lambda_1),a\right>\right)
\\&\times
\sum_{w'\in {\cal W}}\epsilon(w')
\exp\left({2\pi i \over b}\left<w'(\rho^*+\lambda_2),a\right>\right)\,.
\label{degchar1}
\eal
We can now use the Weyl character formulae
\beq
\chi_{R_1}(e^x)= \frac{\sum_{w\in {\cal W}}\epsilon(w)
e^{ \left< w(\rho+\lambda_1),x \right>}}{ \sum_{w\in {\cal W}}\epsilon(w)
e^{ \left< w(\rho),x \right>}}\,,\qquad \chi_{\LR_2}(e^x)= \frac{\sum_{w\in {\cal W}}\epsilon(w)
e^{ \left< w(\rho^*+\lambda_2),x \right>}}{ \sum_{w\in {\cal W}}\epsilon(w)
e^{ \left< w(\rho^*),x \right>}}\,,
\eeq
where $\chi_{R_1}(e^x)$ is the character in representation $R_1$ of $\gfrak$
while $\chi_{\LR_2}(e^x)$ is the character in representation $\LR_2$ of $\Lgfrak$. We therefore 
find that \eqn{degchar1} can be expressed as
\beq
S_{(R_1,\LR_2)\alpha}=\chi_{R_1}\left(e^{2\pi i b a}\right)\chi_{\LR_2}
\left(e^{{2\pi i \over b} a}\right)S_{1 \alpha} \,.
\label{degchar}
\eeq

Finally, for semi-degenerate representations the character is given in (\ref{characsemisum}), where 
each term in the sum corresponds to the non-degenerate character 
(\ref{nongenerate}) evaluated at momentum
\beq
m=\tilde m + bw'(\rho_\ci+\lambda_1)+{1\over b}(\rho_\ci^*+\lambda_2) \,.
\eeq
It is useful to split the sum over $w\in {\cal W}$ 
into a sum over $\tilde w$ in ${\cal W}/{\cal W}_\ci$ and $\tilde w'$ in ${\cal W}_\ci$. 
The action of ${\cal W}_\ci$ on $\tilde m$ is trivial, hence we can repeat the above steps to obtain
\begin{align}
S_{(\tilde m,R_1,\LR_2)\alpha}=i^{K-1} {\sqrt{\hbox{det}(C)}\over |{\cal W}|}
& \sum_{\tilde w\in {\cal W}/{\cal W}_\ci} 
\exp\left(2\pi i \langle \tilde w(\tilde m),a\rangle\right) \chi_{R_1}
\left(e^{2\pi i b \tilde w^{-1}(a)}\right)\chi_{\LR_2}
\left(e^{{2\pi i \over b} \tilde w^{-1}(a)}\right)\notag \\ & 
\prod_{e\in \Delta^+_\ci}4\sin
\bigg(\pi b \langle a,\tilde w(e)\rangle\bigg)
\sin\bigg(-{\pi \over b} \langle a,\tilde w(e^*)\rangle\bigg)\,.
\end{align}

Equipped with these formulae, we can write down the explicit expressions 
for the topological defects in Toda CFTs (we recall that $\mu=Q+m$):
\begin{itemize}
\item Degenerate topological defects ($m=-b(\rho+\lambda_1)-{1\over b}(\rho^*+\lambda_2)$):
\beq
{\cal O}_{(R_1,\LR_2)}(p)
=\int d\alpha\, D_{(R_1,\LR_2)}(\alpha) 
\sum_{\{k\},\{l\}} |\alpha,k;\alpha^*,l \rangle\otimes \langle \alpha,k;\alpha^*,l|\,,
\label{ZZb}
\eeq
where 
\beq
D_{(R_1,\LR_2)}(\alpha)=\chi_{R_1}\left(e^{2\pi i b a}\right)\chi_{\LR_2}
\left(e^{{2\pi i \over b} a}\right)\,.
\label{ZZbb}
\eeq 

\item Non-degenerate topological defects ($m$ purely imaginary):
\beq
{\cal O}_{\mu}(p)=\int d\alpha\, D_\mu(\alpha) 
\sum_{\{k\},\{l\}} |\alpha,k;\alpha^*,l \rangle\otimes \langle \alpha,k;\alpha^*,l|\,,
\label{FZZTb}
\eeq
where \beq
D_{\mu}(\alpha)={ \sum_{w\in {\cal W}} 
\exp\left(2 \pi i \langle w(m), a \rangle \right)
\over\prod_{e>0}4\sin\left(\pi b \langle a,e\rangle\right)
\sin\left({\pi \over b} \langle -a,e^*\rangle\right)} \,.
\label{FZZTbb}
\eeq

\item Semi-degenerate topological defects ($m = \tilde m - (\rho_\ci+\lambda_1)b - (\rho^*_\ci+\lambda_2)/b$ with $ \tilde m$ purely imaginary):
\beq
{\cal O}_{\mu,R_1,\LR_2}(p)=\int d\alpha\, D_{\mu,R_1,\LR_2}(\alpha) 
\sum_{\{k\},\{l\}} |\alpha,k;\alpha^*,l \rangle\otimes \langle \alpha,k;\alpha^*,l|\,,
\label{FZZTsb}
\eeq
where \beq
D_{\mu,R_1,\LR_2}(\alpha) = \sum_{w\in {\cal W}/{\cal W}_\ci} { 
\exp\left(2 \pi i \langle w(\tilde{m}), a \rangle \right) \chi_{R_1}\left(e^{2\pi i b w^{-1}(a)}\right)\chi_{\LR_2}
\left(e^{{2\pi i \over b} w^{-1}(a)}\right)
\over\prod_{e\in \Delta^+-w(\Delta^+_\ci)}4\sin\left(\pi b \langle a,e\rangle\right)
\sin\left(-{\pi \over b} \langle a,e^*\rangle\right)} \,.
\label{FZZTsbb}
\eeq

\end{itemize}

Therefore, topological defects in Toda theory 
insert the expressions (\ref{ZZbb})(\ref{FZZTbb})(\ref{FZZTsbb})
into Toda CFT correlation functions, depending on the choice of representation for the 
topological line operator. These simple insertions arise for any 
pants decomposition of $C_{g.n}$ for which the closed curve $p$
on which the topological defect operator is supported corresponds 
to an internal edge on the associated trivalent graph $\Gamma_\sigma$.

\subsection{Generalized Topological Line Operators}
\label{sec:topjanus}

There is a useful generalization of the notion of a topological
line operator: line operators which effectively change the complex
structure of the Riemann surface they are inserted in. A CFT correlation function 
on a Riemann surface in the presence of such lines operators compute the CFT correlation 
function on the same Riemann surface but with a different complex structure.

Recall that the 
complex structure moduli of a Riemann surface can be identified 
with the length and twist coordinates associated to the $3g-3+n$ tubes that 
comprise a pants decomposition of the surface from $2g-2+n$ trinions. The moduli arise 
from the transition functions between local coordinate patches $z$ and $z'$ near two 
punctures as $z=q/z'$. A different local coordinate system 
given by the conformal transformation $z\to f(z)$ will give a surface 
with a different complex structure. 

Consider a
non-self-intersecting path $p$ on the Riemann surface, and thicken
it to a strip. We assume that under the locally defined conformal
transformation $z \to f(z)$ the image and
counterimage of $p$ under $f$ belong to the strip. Then we
can define line operators such that the energy momentum $T$ on
one side of the line operator equals $f \circ T$ on the other
side of the line operator, and analogously identify $\bar{T}$ and $\bar{f} \circ \bar{T}$.
The symbol $f \circ T$ denotes the action of the conformal
transformation $f$ on the energy momentum tensor $T$
\beq
f \circ T(z)={1\over \left(\partial_z f(z)\right)^2}T(z)-{c\over 12} \{f(z),z\}\,,
\eeq
where $\{f(z),z\}$ denotes the Schwarzian derivative.

Such a generalized line operator can be written as
\beq
{\cal O}_{\mu,f}(p)=\sum_{\alpha\in \ci} 
D_{\mu \alpha} \sum_{\{k\},\{l\}} |f \circ \alpha,k;\bar{f} \circ \alpha^*,l \rangle\otimes \langle \alpha,k; \alpha^*,l|\,,
\label{projectorsa} 
\eeq 
where $|f \circ \alpha, \bar{f} \circ \alpha^*\rangle\!\rangle\equiv\sum_{\{k\},\{l\}} 
|f \circ \alpha,k;\bar{f} \circ \alpha^*,l \rangle\otimes |\alpha,l;\alpha^*,k \rangle$ 
are the permutation Ishibashi states which
implement the twisted gluing conditions across the curve $p$
\beq
f \circ T_1=T_2\qquad \bar{f} \circ \bar{T}_1=\bar{T}_2\,.
\eeq

The coefficient $D_{\mu \alpha}$ here is the same as for the standard topological line 
operators (\ref{modular}). Even if $D_{\mu \alpha}$ is trivial, the
insertion of this line operator in a CFT correlation function
modifies the holomorphic and antiholomorphic conformal blocks
such that they correspond to a Riemann surface with a different complex structure. 

Notice that it is possible to choose functions $f, \bar f$
which are {\it not} related by complex conjugation. 
This requires a little bit of care: the action of $f, \bar f$ on primary fields of generic 
conformal dimension is ill defined because of choice of branch in the Jacobian 
$(\partial f)^{\Delta(\alpha)} (\bar \partial \bar f)^{\Delta(\alpha^*)}$. We really need 
to specify the path we follow in deforming $\bar f$ away from the conjugate to $f$. 
This fixes the choice of branch, but forces us to think in terms of Teichm\"uller space 
rather than complex structure moduli space. 

Inserting such a generalized loop operator in a CFT
correlation function results in a correlation function which pairs holomorphic and
antiholomorphic conformal blocks defined at different values of
the (Teichm\"uller) complex structure moduli. The expression corresponding to the 
insertion of ${\cal O}_{\mu,f}(p)$ for the identity representation is given by
\beq
\langle {\cal O}_{1,\{f,\bar f\}}(p)\rangle_{C_{g,n}}=
\int
d\nu(\alpha)\,{\overline
\cF^{(\sigma)}_{\alpha,E}}(q)\,\cF^{(\sigma)}_{\alpha,E}(q')\,,
\label{Z-Janus}
\eeq 
where $q$ and $q'$ denote two points in Teichm\"uller space.
This pairing makes sense because the space
of conformal blocks is a flat bundle over Teichm\"uller space,
so that conformal blocks at different values of the complex structure can be paired up 
as long as a path is specified between the two choices of complex structure.

Although the deformation of the Rieman surface complex
structure moduli through generalized line operators is only local in
Teichm\"uller space, the pairing between holomorphic and
anti-holomorphic blocks defined at different values of the
complex structure moduli in \eqn{Z-Janus} makes sense, by analytical 
continuation, also for values of the moduli which are arbitrarily far in 
Teichm\"uller space. There is one aspect about eqn. \eqn{Z-Janus} which is 
somewhat unnatural, that the same pants decomposition is used in both 
conformal blocks, even if they describe the Riemann surface at very 
different points in Teichm\"uller space.

It is more natural at a given point to use the pants decomposition along 
``short'' cycles, so if the trivalent graph $\Gamma_\sigma$ is associated with
a natural pants decomposition $\sigma$ at $q$ we can choose the graph 
$\Gamma_{\sigma'}$ for a natural pants decomposition $\sigma'$ at $q'$. 
The two graphs are related to each-other 
by a sequence of steps in the Moore-Seiberg groupoid ${\cal G}$, which 
describes the evolution from a trivalent graph $\Gamma_{\sigma}$ 
to the trivalent graph $\Gamma_{\sigma'}$ mirroring the precise 
path $g^{(\sigma',\sigma)}$ from the point $q$ to $q'$ in Teichm\"uller space. 

The analytic continuation of the holomorphic conformal block is then 
captured by a representation of the 
element of the Moore-Seiberg groupoid ${\cal G}$ as a matrix 
$g^{(\sigma',\sigma)}$ acting on the holomorphic conformal blocks.
The result can be written as 
\beq 
\int
d\nu(\alpha)d\nu(\alpha')\,{\overline\cF^{(\sigma)}_{\alpha,E}}(q)
\,g^{(\sigma,\sigma')}_{(\alpha,\alpha')}\,
\cF^{(\sigma')}_{\alpha',E}(q')\,. 
\label{Z-dwall}
\eeq
A special case is when $q$ and $q'$ are image points under the mapping class 
group, i.e., they map to the same point in moduli 
space. Then we can take $\sigma=\sigma'$ and 
$\cF^{(\sigma')}_{\alpha',E}(q')=\cF^{(\sigma)}_{\alpha',E}(q)$ 
and the above expression would 
differ from the usual correlation function by the inclusion of the 
matrix $g$.

There is a second way to generalize the notion of topological 
line operator, which will not play a role in this paper, but is relevant in matching 
four dimensional ${\cal N}=2$ gauge theories with 2d CFT correlation functions. It is 
possible to twist the definition of a topological line operator by an outer automorphism 
of the chiral algebra ${\cal A}$, so that the chiral algebra currents on the two sides of 
the line operator are related by the action of the outer automorphism. In the case of 
Toda theories, important outer automorphisms are induced by the symmetries of the 
corresponding Dynkin diagram. The theory of twisted line operators is closely related 
to the theory of orbifold constructions.

In the rest of the paper we will advance the four dimensional ${\cal N}=2$ gauge 
theory interpretation of the topological defect operators in Toda CFTs we have 
constructed. We will find that they admit a rather elegant description in terms of 
loop operators and domain walls in gauge theory.

\section{Degenerate Topological Defects in Gauge Theory}
\label{sec:ZZ-loops}

We would like now to apply our results on topological defects in Liouville 
and Toda theories to the study of observables in four dimensional $\cN=2$ gauge theories. 
The relevant theories are denoted ${\cal T}_{g,n}(\gfrak)$ 
\cite{Gaiotto:2009we}, and are associated to a Riemann surface 
$C_{g,n}$. The partition function of these theories on $S^4$ \cite{AGT} as well as certain 
interesting observables \cite{Alday:2009fs,Drukker:2009id} may be calculated 
by Liouville/Toda CFT on $C_{g,n}$. In particular, the partition function of ${\cal T}_{g,n}(\gfrak)$ on $S^4$ conjecturally captures the correlation function of Liouville/Toda CFT on $C_{g,n}$:
\beq
{\cal Z}_{{\cal T}_{g,n}(\gfrak)}\equiv\int[da]\,
\overline{Z}_\text{Nekrasov}\, Z_\text{Nekrasov}
=\Big<\prod_{i=1}^n V_{m_i}\Big>_{C_{g,n}}\equiv\int d\nu(\alpha)\,
{\overline \cF^{(\sigma)}_{\alpha,E}}\,\cF^{(\sigma)}_{\alpha,E}\,.
\eeq
In this section we provide the gauge theory interpretation of the topological 
defect operators corresponding to fully degenerate 
representations in these CFTs.

\subsubsection*{Liouville CFT}

The Wilson and 't~Hooft loop operators in the 
4d ${\cal N}=2$ theories ${\cal T}_{g,n}(A_1)$ were identified in 
\cite{Alday:2009fs,Drukker:2009id} with the Verlinde loop operators 
${\cal L}_\mu(p)$ obtained by transporting the degenerate 
fields of Liouville theory around closed curves\footnote{Homotopy 
classes of curves to be precise.} on the 
Riemann surface $C_{g,n}$. Given a choice of duality frame for 
${\cal T}_{g,n}(A_1)$ or equivalently a choice of pants decomposition 
of $C_{g,n}$, the electric and magnetic charges of the gauge theory 
loop operator associated to a curve $p$ in $C_{g,n}$
are given by the Dehn-Thurston parameters $(p_i,q_i)_{i=1,\ldots,3g-3+n}$ 
of the curve.\footnote{These gauge theory loop operators are supported 
on a fixed equatotial $S^1\subset S^4$. Changing the curve on the Riemann 
surface corresponds to changing the charges of the loop operator in gauge theory.} 
The Dehn-Thurston parameters depend on the choice of pants decomposition 
just as the electric and magnetic charges of a loop operator in the 4d gauge 
theory ${\cal T}_{g,n}(A_1)$ depend on the choice of duality 
frame \cite{Drukker:2009tz}.

We recall that the identification between the 4d ${\cal N}=2$ theories 
${\cal T}_{g,n}(A_1)$ on $S^4$ and Liouville CFT on $C_{g,n}$ 
holds for the value $b=1$, corresponding to Liouville theory with central charge 
$c=25$. At $b=1$, 
the degenerate fields depend only on $r+s$, as can be seen 
by looking at their Liouville momentum (\ref{momdege}). The degenerate 
fields labeled by $(1,s)$ serve as a basis of such operators.

The proposal that the Verlinde loop operators correspond to Wilson and 't~Hooft loop 
operators was explicitly verified by computing the expectation value of the 
Verlinde loop operator along a curve $p$ wrapping around an internal edge in the 
graph $\Gamma_\sigma$ corresponding to a pants decomposition 
of $C_{g,n}$ in Liouville theory. For this class of curves the corresponding gauge 
theory operators are the Wilson loop operators. This was then 
compared with Pestun's exact formula \cite{Pestun:2007rz}%
\footnote{We will review aspects of Pestun's localization in 
Section~\ref{pestun} in order to enrich the dictionary between 2d Toda CFTs and 4d 
gauge theories to include gauge theory domain walls.}
for the expectation 
value of a Wilson loop operator in the 4d ${\cal N}=2$ gauge theory
${\cal T}_{g,n}(A_1)$ on $S^4$.

For $\gfrak=A_1$, a 
Wilson loop operator for one of the gauge groups is labeled by its $SU(2)$ spin $j$, and we denote it by 
$W_j$. 
Explicit computation of the monodromy acquired by the $V_{1,2j+1}(z)$ 
degenerate field defining the Liouville theory Verlinde loop operator 
using the CFT fusion matrices confirmed that \cite{Alday:2009fs,Drukker:2009id}
\beq
\vev{W_j}_{{\cal T}_{g,n}(A_1)}=\vev {{\cal L}_j(p)} _{C_{g,n}}\,.
\label{proposal}
\eeq
Since the explicit form of the fusion matrices for the non-minimal ($j>1/2$) 
degenerate fields is rather complicated to compute, the proposal could only 
be explicitly checked up to $j\leq 5/2$.

Using the identification between Verlinde loop operators ${\cal L}_\mu(p)$ 
and topological defect operators ${\cal O}_\mu(p)$ proven in 
Section~\ref{sec:topdefects} and the 
explicit expressions for the Liouville topological defect operators 
described in Section~\ref{sec:topliouville}, we can now demonstrate 
that the proposal in \cite{Alday:2009fs,Drukker:2009id} is valid for an 
arbitrary representation $j$ of the corresponding Wilson loop. We first note that 
for $b=1$ the 
topological defect operator ${\cal O}_{1,2j+1}(p)$ corresponding to the 
degenerate field $V_{1,2j+1}(z)$ given in (\ref{ZZa}) inserts into the 
CFT correlator on $C_{g,n}$ precisely the $SU(2)$ character in the 
spin $j$ representation%
\footnote{Unlike in \cite{Drukker:2009id}, the normalization we employ 
is such that the Wilson loop is not divided by the dimension of the 
representation. See footnote~\ref{foot:normalization}}
\beq
{S_{(1,2j+1)\alpha}\over S_{11}}=\frac{\sin(2\pi a (2j+1))}{ \sin(2\pi a)}
=\sum_{m=-j}^j e^{4\pi i ma}\equiv\Tr_j e^{2\pi i a}\,.
\eeq
This result combined with (\ref{proof})
\beq
{\cal L}_j(p)={\cal O}_{1,2j+1}\,,
\eeq
confirms the proposal (\ref{proposal}) put forward in \cite{Alday:2009fs,Drukker:2009id} 
for arbitrary spin $j$:
\beq
\vev{ {\cal O}_{1,2j+1}(p)}_{{C}_{g,n}}=\int[da]\, \hbox{Tr}_j\,e^{2\pi i a}\,
\overline{Z}_\text{Nekrasov}\, Z_\text{Nekrasov}\,.
\label{pestunnnnn}
\eeq
This shows that loop operators in 4d ${\cal N}=2$ gauge theories 
${\cal T}_{g,n}(A_1)$ are precisely captured by the topological loop operators in 
Liouville theory associated with degenerate representations.

By choosing loop operators supported on curves $p$ in $C_{g,n}$ traversing 
edges in the trivalent graph $\Gamma_\sigma$ 
(those curves have Dehn-Thurston parameters $p_i\neq 0$) 
one can calculate using the formalism of \cite{Alday:2009fs,Drukker:2009id} the 
vacuum expectation value of 't~Hooft operators.
Liouville CFT was used in \cite{Alday:2009fs,Drukker:2009id} to calculate exactly 
't~Hooft operators in certain 4d ${\cal N}=2$ gauge theories, extending previous 
computations of these observables in ${\cal N}=4$ super Yang-Mills 
\cite{Gomis:2009ir}.

\subsubsection*{Toda CFTs}

For $\cN=2$ theories ${\cal T}_{g,n}(\gfrak)$ with $\gfrak\neq A_1$ the 
complete characterization of Wilson-'t~Hooft operators and their corresponding description 
in Toda CFTs is not known. Clutching this correspondence requires, in particular, 
a generalization of Dehn's theorem classifying homotopy classes of non-self-intersecting 
curves on a Riemann surface $C_{g,n}$. In our discussion of topological web operators 
in 2d CFTs, we observed that such webs can be constructed by combining oriented 
lines joining at trivalent junctions on the Riemann surface. This suggests that the 
generalization of Dehn's theorem involves classifying homotopy classes of oriented 
graphs constructed out of trivalent vertices on a Riemann surface $C_{g,n}$. Even 
though we do not know the complete correspondence, 
we will now identify 
the dual of a Wilson loop operator in ${\cal T}_{g,n}(\gfrak)$ with topological line 
operators in $\gfrak$ Toda CFT on $C_{g,n}$ associated with completely degenerate representations.

The completely degenerate representations of Toda CFT carry momentum 
(\ref{completely}), which is the sum of two highest weights. 
For $b=1$, where the relation between the 4d gauge theory ${\cal T}_{g,n}(\gfrak)$ on $S^4$ and the 
Toda CFT is expected to hold, a basis 
for the degenerate representations is labeled by a single representation $R$ of ${\gfrak}$. 
We now show that a topological defect operator (or Verlinde loop operator) 
based on such a representation and wrapping an internal edge in the graph $\Gamma_\sigma$ 
exactly calculates the Wilson loop in representation $R$ of the gauge group 
associated to that edge.

The topological defect operator ${\cal O}_{(1,R)}(p)$ encircling an edge carrying 
a representation $\alpha$ inserts into the Toda CFT correlation function (\ref{ZZb})
\beq
\chi_{R}\left(e^{2\pi i a}\right)=\frac{\sum_{w\in {\cal W}}\epsilon(w)
e^{ 2\pi i \langle w(\rho+\lambda), a \rangle}}{ \sum_{w\in {\cal W}}\epsilon(w)
e^{ 2\pi i \langle w(\rho),a \rangle}}\equiv \hbox{Tr}_R\, e^{2\pi i a}\,.
\eeq
This is precisely the insertion of a character in representation $R$ of 
$\gfrak$. This exactly captures Pestun's result 
\cite{Pestun:2007rz} for the expectation value of a Wilson loop in 
representation $R$ in the corresponding 4d ${\cal N}=2$ theory 
${\cal T}_{g,n}(\gfrak)$ (assuming the instanton partition functions are indeed 
equal to the conformal blocks):
\beq
\vev{ {\cal O}_{(1,R)}(p)}_{{C}_{g,n}}=\int[da]\, \hbox{Tr}_R\,e^{2\pi i a}\,
\overline{Z}_\text{Nekrasov}\, Z_\text{Nekrasov}\,.
\eeq

This maps the computation of loop operators in 
4d gauge theories ${\cal T}_{g,n}(\gfrak)$ to correlation functions in 
Toda CFTs in the presence of topological defect operators.

\section{Domain Walls, Topological Defects and Localization} 
\label{pestun} 

In this section we compute the partition function of 4d ${\cal N}=2$ gauge theories in $S^4$ in the 
presence of domain walls (three dimensional defects). These include Janus domain walls, 
symmetry breaking domain walls and duality walls. We then show that our results match 
the effect of inserting the topological defect operators in Liouville/Toda CFTs discussed 
in Section~\ref{sec:topliouville}. This section combined with the previous one advances 
a 4d gauge theory interpretation for the topological defects in Liouville/Toda CFTs.

We begin by reviewing the basic ingredients of Pestun's localization \cite{Pestun:2007rz} 
required to compute the partition function 
of 4d ${\cal N}=2$ gauge theories in the presence of domain walls. 
The main idea behind Pestun's localization of correlation functions of 4d $\cN=2$ theories 
on $S^4$ is simple: as long as all the observables in the correlation function preserve 
a common superconformal symmetry $Q$, one can add a $Q$-exact term to the 
Lagrangian without changing the result. If the $Q$-exact term is chosen carefully, 
it can freeze out most degrees of freedom, and reduce the calculation to an integral over 
the surviving modes, with a measure determined by the one-loop determinant 
of the frozen fluctuations. 

Pestun's choice of localization term forces all fields to be constant 
away from the north and south poles of the $S^4$. In the neighbourhood of the poles 
the localization reduces to Nekrasov's localization \cite{Nekrasov:2002qd,Nekrasov:2003rj} 
of the gauge theory in $\bR^4$. 
The key technical points which had to be carefully accounted for in the calculation were the 
requirement of an off-shell formulation of the 
supersymmetry $Q$, a careful gauge-fixing and the precise computation of the one-loop determinants.

Pestun's choice of localization term for the $\cN=2$ vectormultiplet freezes (away from the 
poles) the gauge fields and the real part of the adjoint scalar field. The zeromode $a$ of the 
imaginary part of the scalar field survives. The off-shell formulation 
for the $\cN=2$ vectormultiplets is the familiar one with three auxiliary fields in a triplet of 
$SU(2)_R$ $R$-symmetry. One of the three auxiliary fields $D$ is fixed to be equal to the 
zero mode $ia$ by localization. This will be important for us later. The choice of $Q$ to 
use in the localization breaks $U(1)_R$ completely, and $SU(2)_R$ to an $SO(2)_R$ 
subgroup. This selects which component of the complex adjoint scalar field and 
which auxiliary field can be non-zero. 

We can be a bit more specific: there is an $OSp(2|4)$ subgroup of the $SU(2,2|2)$ 
superconformal group\footnote{We are being a bit imprecise with the reality conditions 
for supergroups in Euclidean signature.} which is the symmetry group of mass-deformed $\cN=2$
gauge theories defined on $S^4$ which includes the rotation group of $S^4$, the above-mentioned 
$SO(2)_R$ and $8$ supercharges. One can further restrict to an $OSp(2|2)$ 
subgroup which includes only rotations which fix the poles and are chiral at the north pole 
and can be realized off-shell on the hypermultiplets. Finally, we can select a specific rotation 
generator $J$, and a supercharge $Q$ which squares to $J+R$, where $R$ 
is the $SO(2)_R$ generator. 

There are many observables which can be added to the correlation function of the 4d 
${\cal N}=2$ gauge theory which preserve $Q$. A nice set of examples are half BPS 
defects, which preserve subgroups of $SU(2,2|2)$ which are defined as the 
fixed point of some involution $\sigma$ of $SU(2,2|2)$. 
They will preserve $Q$ if $Q$ (and hence $J+R$) is fixed by the involution. 
Pestun's original computation focused on supersymmetric Wilson loop operators wrapping 
an orbit of $J$. These Wilson loops are constructed with the same 
component of the adjoint scalar which retains the zero-mode $a$ 
upon localization. The same symmetries ($Q$ and $J+R$) can be preserved by 
supersymmetric Wilson-'t~Hooft loops \cite{Drukker:2009tz,Alday:2009fs,Drukker:2009id}, 
by surface operators \cite{Alday:2009fs} and by domain walls. 
We describe the appropriate involutions in 
Appendix~\ref{sec:susy}. It would be interesting to understand Pestun's localization in 
the presence of these defect operators. In this paper we will consider localization in the presence of   domain 
walls wrapping the $S^3$ equator of $S^4$.

\subsubsection*{Janus Domain Walls }

Janus domain walls \cite{Bak:2003jk,Clark:2004sb}
are the simplest: they are defined as some deformation of the 4d $\cN=2$ 
gauge theory Lagrangian which makes the complexified gauge coupling 
$\tau=\theta/2\pi+4\pi i/g^2_{YM}$ jump across the 
domain wall while preserving 3d superconformal invariance (they are usually 
described as the limit of a smooth profile for the gauge coupling). The Lagrangian 
deformation for the 4d $\cN=2$ gauge theory Janus is presented in 
Appendix~\ref{sec:janus}. It is important for us to note that a jump in $g_{YM}$ introduces 
no extra terms involving $a$ or the auxiliary field $D$ at the wall. 
A jump in $\theta_{YM}$ leads to an extra integral of the instanton density over half 
the space. This integral is equal (as can be seen by integrating by parts) to 
a 3d $\cN=2$ supersymmetric Chern-Simons action, with level
$k=\frac{\Delta \theta_{YM}}{2\pi}$. The fields entering in the Chern-Simons action 
are the gauge field, the scalar $a$ and the auxiliary field $D$ restricted to the domain wall.
Furthermore $a$ is continuous across the wall.

Upon localization, the classical Lagrangian $2 a^2 - D^2$ evaluated on the localization 
locus integrated over the four-sphere gives rise to 
\beq 
\exp \left( \frac{8 \pi^2}{g_{YM}^2} \langle a, a \rangle \right)= (q \bar q)^{- \frac{1}{2} \langle a, a \rangle}\,.
\eeq 
In the presence of a supersymmetric Janus domain wall, the classical action is instead given by 
\beq
\exp \left[\frac{4 \pi^2}{g_{YM}^2} \langle a, a \rangle +\frac{4 \pi^2}{(g'_{YM})^2} \langle a, a \rangle 
- (\theta - \theta') \frac{i \langle a, a \rangle}{2}\right]
= (q \bar q')^{- \frac{1}{2} \langle a, a \rangle}\,.
\eeq
The first two terms are the integrals of the bulk action on each hemisphere, the third term comes 
from the ${\cal N}=2$ Chern-Simons Lagrangian coupling $a D$ (\ref{N=2CS}). 
Furthermore, the instanton contributions localized
at the north pole are computed with instanton factor $q$, while the ones localized at the south 
pole with $\bar q'$. Finally, the one-loop determinant is completely unaffected: it only depends 
on the localization term.

Therefore, the complete partition function of a supersymmetric Janus domain wall in a 
4d $\cN=2$ gauge theory takes the form 
\beq 
\int[da]\, 
\overline{Z}_\text{Nekrasov}(q)\, Z_\text{Nekrasov}(q')\,.
\label{pestunnnnnn}
\eeq

Using the identification between $Z_\text{Nekrasov}$ and Liouville/Toda CFT 
conformal blocks $\cF^{(\sigma)}_{\alpha,E}$, we conclude that the partition function of 
the 4d ${\cal N}=2$ gauge theories ${\cal T}_{g,n}(\gfrak)$ in the presence of a Janus 
domain wall (\ref{pestunnnnnn}) exactly reproduces the insertion of the generalized 
topological line operator ${\cal O}_{1,\{f,\bar f\}}(p)$ (\ref{Z-Janus}) into the 2d Liouville/Toda 
CFT correlation function, where $p$ encircles a tube of the pair of pants decomposition 
of the Riemann surface corresponding to this gauge group.

This construction can be extended further by adding a supersymmetric Wilson loop in the 
representation $R$ of the gauge group on the $S^3$ domain wall. Since the intersection 
of the symmetries preserved by the domain wall and the Wilson loop contains Pestun's 
supercharge $Q$, the exact partition function of the Janus domain wall with a Wilson 
loop at the equator is given by 
\beq 
\int[da]\, \hbox{Tr}_R\,e^{2\pi i a}\,
\overline{Z}_\text{Nekrasov}(q)\, Z_\text{Nekrasov}(q')\,.
\eeq
This precisely reproduces the insertion of the generalized topological operator 
${\cal O}_{(1,R),\{f,\bar f\}}(p)$ (\ref{projectorsa}) into the 2d Liouville/Toda CFT correlation function.

\subsubsection*{Symmetry Breaking Walls}

A second class of gauge theory domain walls which we can identify with topological 
loop operators in Liouville/Toda CFTs were first defined for $\cN=4$ SYM in 
\cite{Gaiotto:2008sa}, and can be generalized to 4d $\cN=2$ theories: these 
are domain walls at which the gauge group $G$ is reduced to a subgroup $H$. 
The basic idea is to set to zero at the domain wall the gauge fields in $\gfrak - \hfrak$. 
To preserve $\cN=2$ supersymmetry in three dimensions, one should also set to 
zero their superpartners in the 3d $\cN=2$ supermultiplet. These include again 
the scalar $a$ and the auxiliary field $D$ in $\gfrak - \hfrak$ 
(see Appendix~\ref{sec:Dirichlet}). 
We will focus on the case where $H$ has full rank. 

The usual Vandermonde determinant $\prod_{e>0} \langle a,e \rangle^2$ 
(where $e>0$ denote the positive roots of $\gfrak$)
included in the measure $[da]$ in Pestun's 
formula is naturally combined with part of the one-loop vectormultiplet 
determinant to the measure $\prod_{e>0} \sin^2 \pi \langle a,e \rangle$. 
This is a natural measure for conformal blocks in Toda CFTs 
(see \cite{Ponsot:1999uf,Drukker:2009id} for the case of Liouville). The most immediate 
effect of the symmetry breaking domain wall in the localization computation 
of the partition function is that the 
integral over the zeromode $a\in \gfrak$ is restricted to $a\in \hfrak$. Hence the 
Vandermonde determinant appearing in the measure $[da]$ is reduced from 
$\prod_{e>0} \langle a,e \rangle^2$ to 
$\prod_{e>0}^{e\in \hfrak} \langle a,e \rangle^2$. 

We claim that the one-loop determinant in the presence of the symmetry breaking 
domain wall combines similarly with the reduced Vandermonde determinant 
$\prod_{e>0}^{e\in \hfrak} \langle a,e \rangle^2$ to give 
$\prod_{e>0}^{e\in \hfrak } \sin^2 \pi \langle a,e \rangle$. 
To see this, we note that such factors arise in the one-loop determinant for 
3d $\cN=2$ gauge multiplets on $S^3$  computed in \cite{Kapustin:2009kz} 
in a similar context of equivariant localization, based on the same isometries 
as for the $S^4$ computation we have here. 
Indeed, the effect of the domain wall should be to remove exactly such one-loop 
determinant factors for the 3d $\cN=2$ gauge multiplets on $S^3$ for the fields 
in $\gfrak - \hfrak$. 
The final result for the partition function of the symmetry 
breaking domain wall is (written in the CFT language)\footnote{We recall that $\alpha=Q+a$.}
\beq
\int
d\nu(\alpha)\frac{\prod_{e>0}^{e\in \hfrak } \sin^2 \pi \langle a,e \rangle}
{\prod_{e>0}^{e\in \gfrak } \sin^2 \pi \langle a,e \rangle}\,{\overline
\cF^{(\sigma)}_{\alpha,E}}(q)\,\cF^{(\sigma)}_{\alpha,E}(q) e^{2 \pi i \langle \tilde m,a\rangle}\chi_R(e^{2 \pi i a})\,.
\label{Z-reduced}
\eeq 
Just as in the case of the Janus domain wall, we have also included a supersymmetric 
Wilson loop in a representation $R$ of $H$, which is allowed to live at the domain wall. 
The term $ e^{2 \pi i \langle \tilde m,a\rangle}$ originates from the Fayet-Iliopoulos parameters $\tilde m/\pi$ for 
the $U(1)$ factors localized on the domain wall, which upon localization yields 
$2\pi^2\langle \tilde m/\pi,D\rangle=2\pi i\langle \tilde m,a\rangle$.

We can compare this computation with the result of inserting a semi-degenerate topological defect operator 
around a tube of the pair of pants decomposition of the Riemann surface: 
$\hfrak$ can be taken to be the union of the Cartan subalgebra and of $\gfrak_\ci$
to obtain a beautiful match with the 2d Liouville/Toda CFT computation (compare 
with eqn. (\ref{FZZTsbb})).\footnote{ Notice that inside the integral over $\alpha$ the 
sum over the Weyl group orbits in $D_{\mu,R,1}(\alpha)$ can be removed by the 
trivial reflection $a \to w(a)$.}

\subsubsection*{Duality Walls}

We would also like to understand the gauge theory realization of the expression (\ref{Z-dwall}), 
obtained by the action of a 
Moore-Seiberg groupoid element $g$ on the antiholomorphic conformal block. 
Morally, this should correspond to the result of acting with the corresponding S-duality 
transformation on a single side of a Janus domain wall. This operation led in the $\cN=4$ 
case to the notion of a ``duality wall'', which interpolates in a supersymmetric fashion between 
two S-dual images of the same theory \cite{Gaiotto:2008ak}. 
In that case, the duality wall for a $T$ 
operation (a shift of the $\theta$ angle by $2\pi$) consisted of an extra supersymmetric 
Chern-Simons term in the action, involving the 
restriction of the gauge multiplet to the domain wall. The duality wall for an $S$ operation 
had to allow the interaction of two theories with gauge groups $G$ and $\LG$ at the 
common interface, where $\LG$ is the GNO or Langlands dual group. It admitted a 
description in terms of a 3d ${\cal N}=4$ SCFT with $G \times \LG$ flavor symmetry. 
The two bulk theories had Neumann boundary conditions for the 
gauge multiplet at the wall, which allowed each factor in the flavor group 
$G \times \LG$ to be gauged. 
This is discussed in a bit more detail in Appendix~\ref{sec:S-duality}.

We expect a similar picture to be valid for 4d $\cN=2$ gauge theories.
For example, a duality wall corresponding to the shift of the $\theta$ angle by $2\pi$ 
will clearly be realized by a 3d $\cN=2$ Chern-Simons term, which already appeared 
in the description of the corresponding Janus wall, and we already learned that it will 
contribute an overall factor of $\exp\left(2 \pi i \langle a, a \rangle \right)$. This exactly agrees with 
the phase factor associated with a Dehn twist of a tube on the holomorphic conformal 
block, which inserts a phase $\exp(2\pi i \Delta(\alpha))$, where $\alpha=Q+a$. 

The duality wall associated with a general S-duality groupoid element $g$ should 
correspond to the coupling of a 3d $\cN=2$ SCFT with appropriate flavor symmetry 
groups to the original theory on one side of the domain wall, and the S-dual theory on the other. 
If we perform localization in the presence of such a duality wall, the scalar zeromodes 
$a$ and $a'$ of the theories on the two sides will be free to vary (they have Neumann 
boundary conditions at the wall) and 
the 3d $\cN=2$ SCFT on the equator will contribute a factor $Z_{S^3}(a,a')$: 
the $S^3$ partition function of the 3d $\cN=2$ SCFT in the presence of an expectation value 
for the 3d $\cN=2$ gauge multiplet scalars coupled to the two flavor symmetry groups.
These are, by definition, real mass parameters for the 3d SCFT. 

The tree-level and instanton contributions from the poles are not affected by the 
presence of the wall. The most subtle element of the computation is the one-loop 
factor, as we do not know how the matter hypermultiplets may be coupled to the 
defect. We encode our ignorance in the measure of integration for the zeromodes $a$ and $a'$
\beq
\int
d\tilde \nu(\alpha) d\tilde \nu(\alpha')\,{\overline
\cF^{(g \circ \sigma)}_{\alpha,E}}(q)\, Z_{S^3}(a,a')\,\cF^{(\sigma)}_{\alpha,E}(q)\,.
\label{Z-dualitywall}
\eeq 
Notice that if we ignore the effect of the matter hypermultiplets, 
and assume that the measure of integration arises from the 1-loop factor for a 3d $\cN=2$
gauge multiplet at the $S^3$, we get 
\beq 
d\tilde \nu(\alpha)=\prod_{e>0} \sin^2 \pi \langle a,e \rangle da .
\eeq
This is the natural measure of integration for conformal blocks in Toda theories. 
We are then led to conjecture that the partition function of the 3d $\cN=2$ SCFT 
living at a duality wall labeled by an S-duality groupoid element $g$ equals the integral 
kernel $g^{(\sigma,\sigma')}_{(a,a')}$, which implements the Moore-Seiberg groupoid 
element corresponding to $g$ in eqn. (\ref{Z-dwall}). 

As discussed earlier, the correlation function of topological webs also have the form 
of a kernel between a holomorphic and an antiholomorphic conformal block 
(\ref{oper-O}). We conjecture that such kernels are associated with the 
partition function on $S^4$ of the 4d $\cN=2$ gauge theory coupled to another 
class of 3d $\cN=2$ SCFT with $G\times G$ flavor symmetry living in the equator 
(possibly enriched by adding line operators along the wall).
In the next section we will suggest a construction of such SCFTs from the 
compactification of the 6d $(2,0)$ SCFT on appropriate three manifolds,
which is known to produce 3d theories with $\cN=2$ supersymmetry.
It would be interesting to give a field theory interpretation of the 
algebraic structure produced by the composition of webs ${\cal O} \circ {\cal O}'$ 
and of the structure coefficients in the composition product. 
Two parallel half-BPS domain walls in a four dimensional superconformal theory may be ``composed'' simply by bringing them together into a new half-BPS domain wall. We conjecture that this operation corresponds to the composition of topological webs in the 2d CFT. 

\section{Defects from the 6d Theory and from M-theory}
\label{sec:M}

At this point we have a variety of topological defects in the Liouville/Toda CFTs
with a proposed four dimensional gauge theory interpretation. In this section we would like to 
further support the match between the 2d correlation functions and 4d gauge theories. A natural approach to bridge the gap between 4d and 2d is 
to recall the 6d field theory construction which is at the base of the definition of 
the 4d gauge theories ${\cal T}_{g,n}(\gfrak)$.

The $(2,0)$ 6d SCFTs are strongly interacting field theories, with no known description 
in terms of a Lagrangian or some elementary degrees of freedom. 
They are labeled by a choice of ADE simply laced Lie algebra $\gfrak$. 
We have three main sources of information about them.
If $\gfrak = \su(N)$, they can be engineered from a set of $N$ M5 branes in flat space. 
They have a Coulomb branch of vacua, where the low energy theory 
reduces to a free theory of self-dual three-form field strengths, fermions and scalars.
Compactified on a circle, they have a low energy description as 5d SYM with a gauge 
group $G$ whose Lie algebra is $\gfrak$. 
Although we cannot define objects directly in the full $(2,0)$ theory, 
we should specify the ``image'' of those objects in these three pictures. 

The theory has a useful half-twist, which allows compactification on a Riemann surface 
$C$ of any genus $g$, while preserving ${\cal N}=2$ SUSY in the remaining four 
dimensions. Codimension $2$ defects located at points on $C$ can be added, 
without breaking any further symmetry.%
\footnote{Some other defects can be thrown 
in: twist lines, which are related to the twisted topological defects in Toda theories 
where the W-algebra is subject to an outer automorphism, and corresponding twist 
fields.} 
In the IR, one recovers the ${\cal T}_{g,n}(\gfrak)$ 4d theories. 
In the language of M5 branes, the twisted compactification corresponds to 
wrapping the M5 branes on $C$ in the space $T^*C$, in the presence of 
additional M5 branes wrapping the fibers above points on $C$. 
The classification of these defects was analyzed as the possible types 
of conformal quiver tails \cite{Gaiotto:2008ak,Gaiotto:2009we}. This was matched 
with the allowed singularities of the relevant large $N$ ``bubbling geometries'' 
\cite{Gaiotto-Maldacena} and to certain vertex operators in Liouville/Toda 
CFTs in \cite{AGT, Wyllard:2009hg, Kanno:2009ga}.

Taken together, these studies suggest that the possible codimension two defects 
are labeled by a choice of an embedding $\rho$ of $\su(2)$ in 
$\gfrak$. They support a flavor symmetry, the commutant of $\rho$ in $\gfrak$ 
and admit ${\cal N}=2$ mass deformations which live in the 
Cartan subalgebra of the flavor symmetry algebra. 
A special case, where $\rho$ is trivial, has flavor symmetry $G$. 
The $A_1$ theory has only one type of defect, the trivial embedding, 
with $SU(2)$ flavor symmetry and a single mass parameter. 

The dictionary between ${\cal T}_{g,n}(A_1)$ theories and Liouville 
correlation functions \cite{AGT} associates this defect with a Liouville vertex 
operator, and the Liouville momentum with the mass parameter. 
More generally, for $\su(N)$ the defects are labeled by a partition of 
$N= \sum_i N_i$, where $N_i$ are the dimensions of the $\su(2)$ irreps 
in the decomposition of the fundamental representation 
of $\su(N)$ under the embedding $\rho$. The $N$ eigenvalues%
\footnote{There are $N$ eigenvalues and $N-1$ mass parameters 
because the vanishing trace.}
of the mass matrix are organized in groups of $N_i$ identical values $m_i$. 
For a trivial embedding, where $N_i=1$, there are $N-1$ mass parameters 
matching the non-degenerate vertex operators in Toda field theories as discussed 
in Section~\ref{sec:Toda-reps}. For more general partitions this matches the 
classification of the semi-degenerate vertex operators in Toda CFT 
with $\lambda_i=0$ \cite{Kanno:2009ga}.

Indeed in $\su(N)$ the simple roots $e_i$ can be taken to be vectors
with the $i$-th entry $1$, and $(i+1)$-th entry $-1$. A vector orthogonal to $e_i$ 
must have the $i$-th entry and $(i+1)$-th entry equal. Hence we have a 
correspondence between the partition $N_i$ and a set $\ci$ of simple roots 
$e_i$: $\ci$ consists of groups of $N_i-1$ consecutive simple roots. The 
Lie subalgebra $\gfrak_\ci$ is $\sum \su(N_i)$. The $T_3$ generator of the 
$\su(2)$ embedding appears to concide with the restricted Weyl vector 
$\rho_\ci$ (hence we will use the same symbol $\rho_\ci$ for the Weyl 
vector and for the corresponding $\su(2)$ embedding). 
The flavor Lie algebra is the commutant of $\gfrak_\ci$ in $\gfrak$.
We see a neat correspondence between the data of $\ci$ modulo the Weyl group
action -- which defines semi-degenerate representations in Toda -- and the data of the $\su(2)$ embedding. 

We expect this correspondence to extend to general simply laced ADE Toda theories 
and $(2,0)$ 6d theories. Going in one direction seems easy: from $\ci$ one gets a Lie subalgebra
$\gfrak_\ci$, then an $\su(2)$ embedding $\rho_\ci$ as the sum of the maximal embeddings 
in the simple factors of $\gfrak_\ci$. The set of mass parameters/Toda momenta orthogonal 
to the roots in $\ci$ is also the commutant to the $\su(2)$ embedding $\rho_\ci$.
It seems a bit more laborious to go in the opposite direction, though it presumably entails 
little more work besides a decomposition of the $T_3$ generator of $\rho_\ci$ 
in simple weights. 

\subsection{Surface Operators in 6d}
\label{sec:6d-surface}
Let us review what is known on the codimension~4 defects, 
which can be described loosely as ``M2-brane defects'': if the 6d field theory
is engineered by a set of coinciding M5 branes, the defect arises as the two-dimensional
boundary of a semi-infinite M2 brane ending on the M5-branes. 

The surface operators in the 6d theory are a sort of generalized 
Wilson line operators. In the Coulomb branch they reduce to 6d Wilson surface 
operators, which couple to the three form field strengths the same way as the 
usual Wilson line operators couple to gauge fields. Upon compactification 
on a circle to 5d they can either wrap the circle, giving rise to half BPS Wilson line 
operators, or not wrap it, giving rise to half BPS 't~Hooft surface operators in the 5d 
SYM theory. Basic M2-brane intersections are the simplest example, giving rise to 
fundamental Wilson loops upon compactification. 

\noindent
Once we consider compactification on a Riemann surface $C$, 
the M2-brane intersections lead to the following defects in the 4-dimensional gauge 
theory:
\begin{enumerate}
\item{\bf Surface operators:}
If the two dimensional intersection is entirely in the 4-dimensional space, these are 
surface operators in the gauge theory. The 6d defects are localized at a point on the 
Riemann surface and should have the interpretation of extra vertex operator insertions in 
Liouville/Toda. Indeed in \cite{Alday:2009fs} the simplest M2 brane defects in the 
$A_1$ theory were identified with $(2,1)$ degenerate vertex operators in Liouville 
theory. They were also identified with the simplest surface operators in $SU(2)$ 
gauge theories, as defined by \cite{Gukov-Witten-surface}.

We are not aware of a direct 4d construction of surface operators which would 
correspond to higher degenerate fields in Liouville theory. 
It may be possible to use the more refined approach in Gukov-Witten, 
and define a defect by coupling a 2d sigma model to the 
4d gauge theory. We do not know of a simple choice of sigma model which would fit the bill. 

A parallel reasoning to the Liouville case identifies the simplest M2 brane 
defects in the $A_{N-1}$ theory with $(N,1)$ degenerate vertex operators in Toda 
theory ($N$ represents the fundamental representation) and also 
with the simplest surface operators in $SU(N)$, defined by 
\cite{Gukov-Witten-surface}, which break the gauge symmetry at the surface to 
$U(1) \times SU(N-1)$. One can also consider defects breaking the gauge symmetry 
to $U(1) \times SU(n) \times SU(N-n)$. These will still have a one-dimensional 
space of parameters. It is far from obvious that this space will coincide with $C$,
but it is reasonable to conjecture that they will correspond again to some 
degenerate fields, possibly labeled by the rank $n$ antisymmetric representation of $SU(N)$. 
These surface operators can be described as well by the coupling of the 4d gauge 
theory to a 2d sigma model with target space the Grassmanian
$SU(N)/(U(1) \times SU(n) \times SU(N-n))$. 
The cohomology ring of the Grassmanian is isomorphic 
to the $n$-th exterior power of a $N$ dimensional space. 
It would be nice to verify whether for more general ADE groups the simple surface 
operators with a single $U(1)$ factor in the unbroken gauge 
group do match degenerate fields in Toda. 

\item{\bf Line operators:}
If one direction of the intersection is in the 4-dimensional space and the other direction 
wraps the Riemann surface we get line operators in the gauge theory. These are 
Wilson, 't~Hooft or dyonic loops, and the charges can be read from the homotopy 
class of the curve on the Riemann surface \cite{Drukker:2009tz}. In 
\cite{Alday:2009fs,Drukker:2009id} these objects were also shown to be represented 
by degenerate fields in Liouville, not by extra insertions, but by their monodromies. 
In Section~\ref{sec:ZZ-loops} we showed that these can also be identified with 
ZZ topological defects in Liouville. The statement extends to Toda: 
the Wilson loop operators correspond to fully degenerate topological defects in Toda CFT. 

This is consistent with the reduction of the 6d theory to 5d SYM: the basic step to recover 
a gauge theory interpretation of the 4d theory is to consider a pair of pants decomposition 
of $C$, make the handles long and thin, and reduce the 6d theory on the handles to 5d 
SYM on segments. An M2-brane defect which wraps around a handle descends to a 
fundamental Wilson loop of the 5d SYM theory, and then to a fundamental Wilson 
loop of the gauge groups of the 4d gauge theory. The same should be true for the 
defects which give rise to Wilson loops in other representations of the gauge group. 

\end{enumerate}

\subsection{Codimension~2 Defects and Domain Walls}
\label{sec:4d-in-6d}

If a codimension~2 defect is extended along some cycle on $C$, it will result in a 3d defect in the 4d space.
At this point, we are ready for a simple analogy. As discussed in 
Sections~\ref{sec:topdefects} and~\ref{sec:topliouville} the classification of topological 
line operators matches that of primary vertex operators. We saw in the 
preceding subsection that the {\em same} 6d surface operators, upon reduction to 4d, 
are related to the to a fully degenerate vertex operator, and to the fully degenerate topological 
line operator in Liouville/Toda CFT. 
We know that the non-fully-degenerate Liouville/Toda vertex operators on $C$ 
are associated to M5-brane defects at points in $C$. Can we give a 6d interpretation 
to the relation we proposed between the non-fully-degenerate topological defects in Liouville/Toda CFTs 
and domain-wall defects in the 4d gauge theories?

Indeed, the domain walls which arise from M5 intersection seem to have the correct 
properties to represent the generic topological line operators in Liouville/Toda theory with 
$R_1 = \LR_2 =1$. They carry the same labels. 
We can push the comparison further if we use a certain result about the nature of 
the codimension~2 intersection. Morally, the 6d defect should admit a description 
in terms of some ${\cal N}=2$ theory coupled to the 6d theory, but as the 6d theory 
is strongly interacting this statement makes little sense. 
Instead, if we compactify on a circle both the bulk theory and the 4d defect, the 
resulting 5d SYM theory is IR free and it makes sense to claim that a certain 
3d SCFT lives on the codimension~2 defect.
There is a reasonable guess: the theory living on the 3d defect should 
coincide with the theory $T_\rho(G)$, which was introduced in \cite{Gaiotto:2008ak}. 
This 3d ${\cal N}=4$ SCFT theory has a $G$ 
flavor symmetry on the ``Coulomb branch'' which can be coupled 
to the 5d bulk gauge fields. It also has a flavor symmetry 
on the Higgs branch, which coincides with he flavor symmetry expected at the 
location of the 6d defect (the commutant to $\rho_\ci$ in $G$). 
If the M5-brane defect wraps a curve around a handle we can try to use the same 
argument which we used for the M2-defects and Wilson loops, and argue that the 
domain wall is given by coupling a $T_\rho(G)$ to the 4d gauge group 
corresponding to the handle. 

The conclusion is a bit hasty: to go from the coupling of $T_\rho(G)$ to the 5d YM 
on the segment to a coupling of $T_\rho(G)$ to the 4d $\cN=2$ gauge theory we 
have to do an RG flow, and it is unclear that $\cN=2$ SUSY in 3d will be sufficient to protect us 
from quantum corrections. One the other hand is we turn on the mass parameters $m$ 
for the flavor symmetry of the operator, $T_\rho(G)$ has a good description as a 
sigma model whose target space is the orbit of $m$ under the coadjoint action of 
the complexified $G^C$. Such a sigma model is exactly what one needs to Higgs the $G$ 
gauge symmetry at the domain wall to the subgroup $H$ we encountered in 
Section~\ref{pestun}. We conclude that the 6d construction is consistent with 
our direct 4d analysis. 

\subsection{More 6d constructions}
It would be interesting to understand the meaning of semi-degenerate representations 
with general representation $R_1,\LR_2$. They may correspond to a combination of a codimension 2 
defect and a codimension 4 defect which is allowed to live on the codimension 2 
defect only. In gauge theory, we saw they may correspond to line operators which 
exist only at the domain wall. 

As a possible generalization, 
rather than just adding defects to the 6d theory compactified on $C$, we can consider a setup 
where the compactification scheme itself is modified. A half BPS domain wall in a 4d ${\cal N}=2$
theory preserves 3d ${\cal N}=2$. There is a natural twist of the 6d theory which preserves 
the same amount of SUSY on a manifold of the form $\bR^3 \times M$, where $M$ 
is a generic 3-manifold. It corresponds to wrapping M5 branes on $M$ inside $T^*M$. If 
$M = \bR\times C$, we are back to the 4d theories. If $M$ is a cylindrical manifold, which 
asymptotes to $\bR\times C$ to both ends, we are building some 
domain wall for the 4d theory. A simple example is a Janus domain wall, where $M$ is 
topologically $\bR\times C$ but the complex structure of $C$ (i.e. the gauge couplings) 
evolves from one end to the other. If the two endpoints of $M$ differ in the topology of 
$C$, the result is a domain wall interpolating between different 
gauge theories. Furthermore, the M5 brane defects can be added along any curves in the geometry. 

An extreme case would be a boundary condition: a three manifold which is asymptotically 
$\bR^+ \times C$
but closes off at the other end. Topologically, this is a three-manifold which is ``filling in'' 
some of the handles of the Riemann surface. There is one such manifold for each pair of pants decomposition, 
suggesting that this should be a rather canonical boundary condition on the gauge fields associated 
to that pair of pants decomposition. As the handles of the Riemann surface are getting 
thinner towards the boundary, the gauge couplings are becoming very weak. We conjecture that the 
boundary condition should correspond to decoupling of all gauge groups, i.e. Dirichlet boundary conditions. 
Instead of closing off the handles smoothly, as boundaries of contractible disks, we could insert 
a codimension 2 defect at the center of the disks, which would give a skeleton of line defects 
in the three manifold, in the shape of the corresponding conformal block diagram and 
carrying the same labels. We conjecture that this represents a Dirichlet boundary condition 
with fixed non-zero value of the scalar fields $a_i$.

Dirichlet b.c. are a useful tool in studying duality walls: in order to identify 
which 3d SCFT lives at a domain wall, one can ``turn off'' the bulk gauge fields by 
placing the theory on a segment, with Dirichlet b.c. at the two ends, and the domain wall 
in the middle. If we lift this setup to the 6d theory, we will see a 3 manifold 
built from the three manifolds which represent the Dirichlet b.c. by gluing the boundaries, 
upon the action of the appropriate Moore-Seiberg groupoid transformation. 
The resulting three manifold will contain a network of codimension two defects, 
obtained by gluing together the two conformal block diagrams at the external legs. 

For example, the SCFT associated with the crossing symmetry of a four punctured sphere 
will arise from an $S^3$ containing a neat tetrahedron, whose six sides 
are labeled by the four external labels, and the two internal labels of the initial and final 
conformal blocks. This setup has the symmetries which are expected for the fusion matrix, 
which in the natural normalization coincides with a quantum $6j$ symbol (at least in Liouville theory). 
Another example is the SCFT associated with the $S$ move on a torus:
the three manifold is again an $S^3$, with two circular defects linked together. 
If true, this gives a surprising description of $T(G)$. In this spirit, the SCFT associated to a web $\cO$ would arise from a 
three manifold where the union of the two conformal block diagrams is linked to the
web. 

There is a strong resemblance between the various choices of $M$ and 
the 3d TFT diagrams which can be used to describe/compute conformal blocks in 
2d CFTs. These are some exciting directions in which the constructions in this 
paper can be generalized.

\section*{Acknowledgements}
We would like to thank T. Okuda and J. Teschner for helpful
discussions. A preliminary version of these results were presented by J.G. at 
the Princeton Center for Theoretical Science in October
and at the Arnorld Sommerfeld Center in Munich in December. J.G. would like 
to thank these institutions and the Institute for Advanced Studies at Princeton for 
their hospitality. D.G. would also like to thank the Perimeter Institute for hospitality.
Part of this work was done during the Workshop on Interfaces and 
Wall-crossings at the Ludwig-Maximilians-Universit\"at in 
Munich, and we would like to thank the organizers for their invitation. 
N.D. would like to thank the Weizmann Institute for its hospitality.
D.G. is supported in part by the Roger Dashen membership in the Institute for Advanced 
Study. D.G. is supported in part by the NSF grant PHY-0503584. 
Research at the Perimeter Institute is
supported in part by the Government of Canada through NSERC and by
the Province of Ontario through MRI. J.G. also acknowledges further
support from an NSERC Discovery Grant and from and an ERA grant by the 
Province of Ontario.

\appendix

\section{Gauge Theory Operators and Localization}
\label{sec:susy} 

The defects in 4d gauge theories can have an arbitrary dimension: domain walls, 
surface operators, line operators and local operators. In this appendix we focus on 
maximally supersymmetric defects, which preserve a maximal subgroup of the 
superconformal group $SU(2|2,2)$, or if the theory is not conformal, of the superisometry 
group $OSp(2|4)$ of $S^4$. We focus on the subalgebras that contain the supercharge 
selected by Pestun (together with the associated $SU(1|1)$ symmetry) to perform 
localization of the path integral. These subalgebras can be 
described as the fixed locus of an involution of $SU(2|2,2)$, which extends 
the obvious geometrical involution which fixes the location of the defect 
in $\bR^4$ or $S^4$.

\subsection{Involutions and Subalgebras}
\label{sec:subalgebras} 

Our discussion will switch between flat space with Lorentzian signature,
and Euclidean space and the four sphere, hopefully raising no serious confusions.
In general, a flat $d$-dimensional defect in a flat $D$-dimensional spacetime
will break all translations and special conformal
transformations in directions transverse to the defect. 
It will also break the rotation group to 
$SO(d) \times SO(D-d)$. The subgroup of the conformal group $SO(D,2)$ which 
leaves the location of the defect unchanged is $SO(d,2) \times SO(D-d)$. This is 
the extension to all spacetime of the $d$-dimensional conformal transformations, 
and the rotations of the transverse directions.
It is useful to identify this subgroup as the fixed locus of an involution,
involving the reflection of the coordinates transverse to the defect.
If we extend this involution by an involution on the $R$-symmetry such that it 
preserves Pestun's supercharge, then the subgalgebra in $SU(2|2,2)$ or 
$OSp(2|4)$ fixed by this involution is the minimal symmetry group that 
must be preserved by the defect.

Let us construct the subalgebras preserved by the defects of different dimensionality. 
The supercharges of the $\cN=2$ superalgebra in flat space are
$Q_\alpha^a$ and $\bar Q_{\dot\alpha}^a$. The superconformal
generators are $S_\alpha^a$ and $\bar S_{\dot\alpha}^a$. They satisfy the algebra
\bal
\{Q_\alpha^a,\,\bar Q_{\dot\alpha}^b\}
&=\varepsilon^{ab}\gamma^\mu_{\alpha\dot\alpha}P_\mu\,,
\qquad&
\{Q_\alpha^a,\,S_{\beta}^b\}
&=\varepsilon^{ab}J_{\alpha\beta}
+\varepsilon_{\alpha\beta}R^{ab}
-\frac{1}{2}\varepsilon_{\alpha\beta}\varepsilon^{ab}D\,,\\
\{S_\alpha^a,\,\bar S_{\dot\alpha}^b\}
&=\varepsilon^{ab}\gamma^\mu_{\alpha\dot\alpha}K_\mu\,,
&
\{\bar Q_{\dot\alpha}^a,\,\bar S_{\dot\beta}^b\}
&=\varepsilon^{ab}\bar J_{\dot\alpha\dot\beta}
+\varepsilon_{\dot\alpha\dot\beta}R^{ab}
-\frac{1}{2}\varepsilon_{\dot\alpha\dot\beta}\varepsilon^{ab}D\,.
\eal
We shall used the antisymmetric tensor with the convention $\epsilon^{+-}=-\epsilon_{+-}=1$ 
to raise and lower all indices. 
$J_{\alpha\beta}$ and $\bar J_{\dot\alpha\dot\beta}$ are the chiral and anti-chiral rotations
and $R^{ab}$ are the R-symmetry transformations.

It is useful to group all the supersymmetry transformations together and write the
general supersymmetry transformation as
\beq
\delta_\epsilon X=\big[(\epsilon_0 Q+\epsilon_1 S),\,X\big],
\eeq
where $\epsilon_0$ and $\epsilon_1$ are constant spinors. In flat space $\epsilon_0$
is a Killing spinor while $x^\mu\gamma_\mu\epsilon_1$ is a conformal Killing spinor.
Invariance under the involutions for the different defects is then conveniently expressed 
as projection equations on $\epsilon_0$ and $\epsilon_1$.

The relevant case for us is when the theory is on $S^4$, and not necessarily 
a conformal theory. In this case the superisometries on the $S^4$ generate 
the $OSp(2|4)$ subalgebra of $SU(2|2,2)$. The supercharges for this supergroup are
\beq
\label{S4-supercharges}
q_\alpha^a=Q_\alpha^a+(\sigma^3)^a_{\ b}S_\alpha^b\,,\qquad
\bar q_{\dot\alpha}^a=\bar Q_{\dot\alpha}^a+(\sigma^3)^a_{\ b}\bar S_{\dot\alpha}^b\,,
\eeq
which means that $\epsilon_{1a}=(\sigma^3)_a^{\ b}\,\epsilon_{0b}$. In order 
to perform Pestun localization, one can pick any real combination of the four 
chiral supercharges $q_\alpha^a$, say $q_{-}^1+q^2_{+}$ (in Euclidean 
signature $(q_{-}^1)^\dagger=q^2_{+}$ and $(q_{-}^2)^\dagger=-q^1_{+}$).%
\footnote{Mass terms for the vector multiplet
will lead to theories with $\cN=1$ supersymmetry. The preserved supercharges are
$\left\{Q_\alpha^1\,,\bar Q_{\dot\alpha}^2\,,S_\alpha^2\,,\bar S_{\dot\alpha}^1\right\}$, 
which can also be written as the conditions
$\gamma^5(\sigma^3)_a^{\ b}\,\epsilon_{0b}=\epsilon_{0a}$ and 
$\gamma^5(\sigma^3)_a^{\ b}\,\epsilon_{1b}=-\epsilon_{1a}$. 
These deformations preserve an $SU(2,2|1)$ supergroup, but 
are incompatible with Pestun's localization} 

This supercharge squares to the sum of the R-symmetry generator 
and a rotation. After the stereographic projection to 
flat $\bR^4$, which can be written as $\bC^2$ with coordinates $z_1$ and $z_2$, the 
relevant rotation is the one acting by $z_i\to e^{i\alpha}z_i$. We shall analyze now 
all possible submanifolds invariant under this symmetry and the associated 
superalgebras.

\begin{enumerate}

\item{\bf Domain walls:}
The most symmetric domain wall is an $S^3\subset S^4$. It is given by the
condition $|z_1|^2+|z_2|^2=1$. Under the stereographic projection we get
an $S^3$ of fixed radius, say unity. It is invariant under all rotations
$J_{\mu\nu}$ as well as the four combinations $P_\mu+K_\mu$.
Similarly the supercharges annihilating it involve combinations of
$Q$ and $S$. These are
\beq
Q_\alpha^a+(\sigma^3)^a_{\ b}S_\alpha^b\,,\qquad
\bar Q_{\dot\alpha}^a-(\sigma^3)^a_{\ b}\bar S_{\dot\alpha}^b\,,
\eeq
This can be expressed by the condition $\epsilon_1=\gamma^5\sigma^3\epsilon_0$.
These generate the superalgebra $OSp(2|4)$. When the theory on $S^4$ 
is not conformal, we obtain the $OSp(2|2)\times SU(2)$ subalgebra of $OSp(2|4)$, 
containing the chiral supercharges $q_\alpha^a$. This is the supersymmetry 
algebra of an $\cN=2$ 3d theory on $S^3$.

There are many more possible geometries of domain walls which preserve
the $U(1)$ symmetry required for localization. The vanishing locus of any equation of
$|z_1|$ and $|z_2|$ will be such a hypersurface. Generically they will
preserve only $U(1)\times U(1)$ symmetry, but there are special
cases with extra symmetries. In $\bR^4$ one can take the condition
$|z_2|=R$ which gives the geometry $\bR^2\times S^1\subset\bR^4$
preserved by $P_1$, $P_2$, $J_{12}$ and $J_{34}$. On $S^4$
the condition $|z_2|R =(1+|z_1|^2+|z_2|^2)$ gives $S^2\times S^1\subset S^4$
which is fixed by $J_{12}$, $J_{23}$, $J_{31}$ and $J_{45}$. It would
be interesting to try to find BPS domain walls of these more general
geometries.

\item{\bf Surface operators:}
The condition $z_2=0$ gives a plane through the origin in flat space
(or an $S^2$ through the poles of $S^4$). This preserves the bosonic generators
$P_1$, $P_2$, $K_1$, $K_2$, $J_{12}$, $J_{34}$ and $D$. 

The surface operator will preserve independently some Poincar\'e and conformal
supercharges. It is convenient to label supercharges with respect to the chirality 
operator on the worldvolume of the surface operator. Then half of the four 
dimensional supercharges have positive chirality and half negative chirality. In $\bR^4$, a 
surface operator can have either $(2,2)$ or $(0,4)$ two dimensional supersymmetry, 
but Pestun's choice of supercharge for localization requires a combination of 
supercharges of positive and negative chirality. Therefore, the $(0,4)$ surface 
operators are not compatible with Pestun's localization.

In more detail, we need to impose the conditions
\beq
\epsilon_0=\pm i\gamma^{34}\epsilon_0\,,\qquad
\epsilon_1=\pm i\gamma^{34}\epsilon_1\,.
\eeq
There are two consistent choices of signs involving the extra gamma matrices
$(\sigma^3)^a_{\ b}\gamma^5$. This results in surface operators with either the
same or opposite chiralities in two dimensions. The first case is
\beq
\epsilon_0^a=i\gamma^{34}\epsilon_0^a\,,\qquad
\epsilon_1^a=-i\gamma^{34}\epsilon_1^a\,,
\eeq
which gives the supercharges
\beq
\left\{Q_-^a\,,\bar Q_+^a\,,S_+^a\,,\bar S_-^a\right\}.
\eeq
These close onto the $SU(1,1|2)$ superalgebra, which is the chiral $(0,4)$ superalgebra 
in two dimensions. In addition there is an extra $SU(1,1)$ symmetry preserved 
by these operators. We can see that this set of supercharges does not intersect 
those in \eqn{S4-supercharges}.

The second choice is
\beq
\epsilon_0^a=i\gamma^{34}\gamma^5(\sigma^3)^a_{\ b}\epsilon_0^b\,,\qquad
\epsilon_1^a=i\gamma^{34}\gamma^5(\sigma^3)^a_{\ b}\epsilon_1^b\,.
\eeq
which gives the supercharges
\beq
\left\{Q_-^1\,,\bar Q_+^2\,,S_+^2\,,\bar S_-^1\right\},
\qquad
\left\{Q_+^2\,,\bar Q_-^1\,,S_-^1\,,\bar S_+^2\right\}.
\eeq
Each set closes on a superalgebra, so the preserved supergroups is 
$SU(1,1|1)\times SU(1,1|1)$. The intersection with \eqn{S4-supercharges} 
is $SU(1|1)\times SU(1|1)$.

Again there are more general 2-manifolds which are invariant under the $U(1)$
symmetry, they are given by the vanishing locus of two equations of
$|z_1|$ and $|z_2|$. The solutions to this condition are of the form
$|z_1|=R_1$ and $|z_2|=R_2$ which give $S^1\times S^1$. It would be
interesting to construct BPS surface operators with this geometry.

\item{\bf Line operators:}
We impose $z_2=0$ and $|z_1|=R$. This leads to an $S^1$ along a latitude on
$S^4$ or on $\bR^4$. The bosonic symmetries are generated by
$P_1+K_1$, $P_2+K_2$, $P_3-K_3$, $P_4-K_4$ $J_{12}$ and $J_{34}$.
The supersymmetry parameters should satisfy
\beq
\epsilon_1=i\gamma^{12}\epsilon_0\,.
\eeq
The resulting generators are
\beq
Q_\alpha^a-(\sigma^3)_\alpha^{\ \beta}S_\beta^a\,,\qquad
\bar Q_{\dot\alpha}^a+(\sigma^3)_{\dot\alpha}^{\ \dot\beta}\bar S_{\dot\beta}^a\,,
\eeq
The preserved supergroup is $OSp(4^*|2)$ and its overlap with the supercharges 
in \eqn{S4-supercharges} generate $SU(2|1)$.

\item{\bf Local operators:}
We impose $z_1=z_2=0$. This is the origin in $\bR^4$. On $S^4$ we should
consider a pair of operators at the south and north poles.
The bosonic symmetries are generated by $J_{\mu\nu}$ and $D$.
Primary operators at the origin of $\bR^4$ are also annihilated by $K_\mu$,
but this symmetry is broken by the operators at the north pole.

As for the surface operator, there will be independent conditions on the
Poincar\'e and conformal supercharges
\beq
\epsilon_0=\pm\gamma^5\epsilon_0\,,\qquad
\epsilon_1=\pm\gamma^5\epsilon_1\,.
\eeq
The preserved supercharges have specific chirality in four dimensions.

There
are two independent possibilities, depending on whether the chiralities of the
two supercharges are the same or opposite. The chiral choice are the 
supercharges
\beq
\left\{Q_\alpha^a\,,S_\alpha^a\right\}.
\eeq
The preserved symmetry is $SU(2)\times SU(2|2)$ and includes Pestun's choice of 
supercharge. These are the operators created out of the vectormultiplet scalars, 
whose expectation value parameterizes the Coulomb branch. The non-chiral choice is
\beq
\left\{Q_\alpha^1\,, \bar Q_{\dot\alpha}^1\,,
S_\alpha^2\,, \bar S_{\dot\alpha}^2\right\}.
\eeq
The preserved supergroup is $SU(2|1)\times SU(2|1)$ and is incompatible with 
Pestun's choice of supercharge. These are operators whose expectation value 
parameterizes the Higgs branch. 

\end{enumerate}

\section{Gauge theory domain walls}
\label{sec:glue-paste}
The theory of domain walls and boundary conditions in $\cN=4$ SYM is rather intricate,
but can be usefully organized by a simple strategy. The same strategy works for $\cN=2$ 
theories as well: 
one can ``cut'' the gauge theory at some location, say $x^3=0$, and impose Dirichlet 
boundary conditions on the gauge fields of the theory, in order to 
decouple the dynamics on the two half spaces. The Dirichlet boundary 
conditions set the field strengths in the $x^3=0$ plane 
to zero. In order to preserve at least half of the super(conformal) symmetries one 
should also impose appropriate b.c. on the fermions and scalar fields in the vector multiplet.
The gauge transformations also are restricted to the identity at $x^3=0$. 
As a result, one has a $G$ flavor symmetry acting on the values of the fields at $x^3=0$
(we denote the gauge group of the bulk theory as $G$). 

One can glue back the theory together by inserting at $x^3=0$ a 3d gauge theory 
with $\cN=2$ 3d SUSY coupled to the two $G$ flavor symmetries of the bulk theories. 
If no kinetic terms for the 3d gauge multiplet are present, the standard couplings of the 
3d gauge multiplet to the two $G$ flavor symmetry currents and moment maps will 
reconstruct the original Lagrangian. If we modify the setup before gluing back, 
we can produce defects at $x^3=0$. For example, the complexified gauge coupling 
on the two sides may be different. The resulting defect is known as a Janus domain wall.
We can also add some 3d degrees of freedom at $x^3=0$, say a 3d $\cN=2$ SCFT 
with $G$ flavor symmetry, and couple them in a standard way to the 3d gauge fields 
introduced in the gluing procedure. The result is a coupling of the 4d gauge theory to the 
3d SCFT. Finally, we may consider a 3d SCFT with $G \times G$ flavor symmetry, 
and add two copies of the 3d gauge fields, one coupled to the bulk theory on $x^3<0$ 
and the first flavor group of the 3d theory, the other coupled to the bulk theory on $x^3>0$, 
and to the second flavor group. The latter construction can be clearly generalized to a 
setup where the gauge theories on the two half spaces differ.

\subsection{Dirichlet Boundary Conditions for $\cN=2$ Gauge Theories}
\label{sec:Dirichlet}

Following this logic, before discussing domain walls, 
we would like to describe how to impose half BPS Dirichlet boundary conditions
for $\cN=2$ vectormultiplets and their interplay with Pestun's localization. 
Here we consider the theory on $S^4$ and impose the boundary conditions 
on the $S^3$ equator such that they preserve the 
$OSp(2|2) \times SU(2)$ subalgebra discussed in Appendix~\ref{sec:susy}. 
This is just the 
symmetry algebra of a 3d $\cN=2$ theory on $S^3$.

The values of the bulk fields 
(and their normal derivatives) at the boundary transform in multiplets of 
3d $\cN=2$ supersymmetry as we discuss in detail in the next 
subsection (based on \cite{Erdmenger:2002ex}). 
Dirichlet boundary conditions for the gauge fields (parallel to $S^3$) have 
to be completed to Dirichlet boundary conditions for all the fields in the same 3d 
multiplet. These gauge fields are in the same multiplet as the scalar field 
which gets a zero mode upon localization. A simple way to see that, is to 
compare Pestun's localization formulae with the localization formulae for a 3d 
gauge multiplet on $S^3$ \cite{Kapustin:2009kz}, which employ the same supercharge. 
This localization does not set to zero the constant mode of the real scalar field, 
and sets it to be equal to the value of the auxiliary field in the gauge multiplet. 
This is exactly the condition imposed by Pestun's localization.

Hence the appropriate Dirichlet b.c. fixes the value of the scalar field in the vectormultiplet 
whose zero mode is the integration variable in Pestun's localization formula. 
It is easy to argue (following for example the discussion in $\cN=4$ SYM \cite{Gaiotto:2008sa})
that supersymmetry also forces the other real scalar field in the 4d vectormultiplet 
to have Neumann boundary conditions. We expect that the result of localization will be 
\beq
\cZ_\text{Dirichlet} = \cF_\alpha(q)\,.
\eeq
The proper normalization for the conformal block $ \cF_\alpha(q)$ should be fixed by a 
one-loop calculation, and may depend on the specific boundary condition selected for the 
matter hypermultiplets. 

This description of the supersymmetric boundary conditions extends to 
domain walls which impose Dirichlet boundary conditions 
on a subset of the gauge fields. An example are the ``symmetry breaking walls'' 
discussed in Section~\ref{pestun} which restrict the gauge 
symmetry to a subgroup at the wall and set the corresponding gauge fields to zero there.

\subsection{Janus Domain Wall}
\label{sec:janus}

As a first example of a half BPS domain wall, we describe in this section
the Janus domain wall: a simple deformation of the gauge theory Lagrangian which
makes the gauge couplings different on the two sides of the wall 
\cite{Bak:2003jk,Clark:2004sb}. 
We present here the gauge theory construction which in Section~\ref{pestun} is 
matched with the generalized topological operators constructed in 
Section~\ref{sec:topjanus}. 

If we have a gauge theory, and modify the standard kinetic term 
to $g^{-2}(x^3) F_{\mu \nu} F^{\mu \nu}$ where $g^{-2}(x^3)$ is a step function,
the Bianchi identities and equations of motions require the tangential 
components $F_{ij}$ of the field strength to be continuous across the wall 
($i,j$ indices run over $0,1,2$ here), while the normal field strengths are rescaled 
across the wall, so that $g^{-2}(x^3) F_{3i}$ is continuous across
the wall (unless matter sources are localized at the wall, in which 
case the discontinuity of $g^{-2}(x^3) F_{3i}$
equals the source current at the wall). 

In order to promote this configuration to a supersymmetric domain wall, we want to pick 
boundary conditions for the other vector multiplet fields such that the normal component 
of the supersymmetry current is 
continuous across the wall, for the subset of the 4d super(conformal) charges which may 
be preserved by a domain wall. Under that subset of SUSY trasformations, $F_{ij}$ and 
$F_{3i}$ do not sit in the same supermultiplet anymore. By supersymmetry, all the fields 
which appear in the SUSY transformations of $F_{ij}$ will have to be continuous across 
the wall, while all the fields which appear in the SUSY transformations of $F_{3i}$ will 
have to be continuous after rescaling by $g^{-2}$. Either by referring to the literature 
\cite{Erdmenger:2002ex}, or by splitting the fields into even and odd under the involution 
of the 4d SUSY algebra which selected the half-BPS subspace, we see that half of the 
fermions, one of the real scalar fields in the vector multiplet $a_1$, and the normal 
derivative of the second real scalar field $D_3 a_2$ have to be 
continuous at the interface, while the other half of the fermions, $a_2$ and $D_3 a_1$ 
are continuous only if mutiplied by $g^{-2}(x^3)$. It is easy to verify that the normal 
component of the supercurrent is then continuous as well: the supercurrent has an 
overall $g^{-2}$ factor and is a bilinear of fields with the two different types of boundary 
conditions. In the Euclidean theory on $S^4$ $a_1$ is related to 
the scalar field $a$ by $a = i a_1$. 

It is useful to develop a technology which allows one to write down the most general 
4d Lagrangians compatible with the 3d SUSY algebra preserved by the domain wall. 
This was done in \cite{Erdmenger:2002ex}, by organizing the 4d fields in supermultiplets 
of the 3d SUSY. In a sense, the 4d fields are treated (classically!) as 3d fields valued 
in functions of $x^3$. The group of 3d gauge transformations is also taken to be the 
group of maps from the $x^3$ line to the 4d gauge group. 
The 4d vector multiplet decomposes into two 3d supermultipelts: a linear multiplet 
containing the 3d part of the connection, $a_1$ and an auxiliary field $D$ which will 
be related by the equations of motion to $D_3 a_2$ and a chiral multiplet which 
includes the complex ``scalar'' $A_3 + i a_2$ and has an auxiliary field F 
which will be set by the equations of motion to $D_3 a_1$. 

We can write the usual Lagrangian for a pure 4d ${\cal N}=2$ gauge
theory in this ${\cal N}=2$ 3d language, and it turns out to simply coincide with the sum of 
the kinetic terms for the 3d linear multiplet and for the 3d chiral multiplet.
The chiral multiplet formed by $A_3 +i a_2$
transforms in a peculiar version of the adjoint representation of the 3d gauge group, 
including a derivative in the $x^3$ direction 
\begin{equation}
\delta_U A_3 + i a_2 = [U,A_3 + i a_2] + \partial_3 U
\end{equation}
No gauge invariant superpotential can be written in terms of this chiral field alone. 

The 4d kinetic terms containing covariant derivatives in the $0,1,2$ directions have an 
obvious 3d origin, but the terms involving the covariant derivative in the $3$ direction 
arise in a more subtle manner from the 3d auxiliary fields. Remember 
that, schematically, the bosonic part of the Lagrangian for the 3d gauge multiplet takes the
form
\begin{equation}
\frac{1}{g^2}\int dx^3 \Tr\left(\frac{1}{2} F_{ij}F^{ij} + D_i a_1 D^i a_1 - D^2 \right).
\label{3d-vector}
\end{equation}
Here the indices $i,j$ run over $0,1,2$, $D_i$ are covariant
derivatives and $D$ the auxiliary field.

The kinetic terms for a chiral multiplet with a scalar $q$ and an auxiliary 
field $F$ take the general form
\begin{equation}
\Tr\left(D_i q D^i \bar q + |a_1 q|^2+2 \bar q D q +F \bar F \right).
\label{3d-chiral}
\end{equation}
We suppressed here the gauge indices, the adjoint fields $a_1,D$ are acting 
in the obvious way on the chiral multiplet fields. 

For the case of the chiral multiplet which includes $A_3+ia_2$, 
the 3d Lagrangian \eqn{3d-chiral} takes the form
\begin{equation}
\frac{1}{g^2}\int dx^3 \Tr\left(F_{3i} F^{3i} + 
D_i a_2 D^i a_2 + D_3 a_1 D^3 a_1 + [a_1,a_2]^2 + D D_3 a_2\right).
\label{3d-adjoint}
\end{equation}
We see the mixed kinetic terms for the 4d gauge fields, derivatives in the $3$ 
direction of the scalar field $a_1$ and, integrating out the auxiliary field $D$, 
the $3$ derivatives of $a_2$.
We included a prefactor $\frac{1}{g^2}$, so that together with \eqn{3d-vector} we 
get the full action for the 4d vector multiplet.

Let us formulate again the basic strategy: 
we want a deformation of the Lagrangian of
an ${\cal N}=2$ gauge theory, which makes the
couplings into functions of the $x^3$ direction. Initially, we
can consider a smooth functional dependence, and aim to
preserve ${\cal N}=2$ super Poincare symmetry in three
dimensions. Then we can make the profile of the couplings into
a sharp step function, in order to localize the wall at a fixed
$x^3$ position. In a 4d SCFT, this step can be identified with
an RG flow to the far IR. With a little bit of care, the
resulting wall will preserve the whole 3d ${\cal N}=2$
superconformal group in the sharp limit. Terms in the
Lagrangian which are proportional to the first derivative of
the couplings will simply localize at the wall, as the derivative of a step function 
is a delta function. Terms with a
more complicate coupling dependence, such as higher derivatives, or
higher powers of the derivatives, should be reabsorbed into
field redefinitions or canceled by extra boundary conditions. If that is not possible, they signal an
inconsistency of the sharp limit.

To construct the Janus solution we would like to allow the gauge coupling 
(and theta angle) to depend on $x^3$. The factor of $1/g^2$ was introduced 
in \eqn{3d-adjoint} by hand to match with \eqn{3d-vector}, but since it does not 
commute with $D_3$, we need to be a bit more careful about how the coupling is 
introduced into the chiral multiplet action \eqn{3d-adjoint}. There is a shortcut:
under the rescalings $a_2\to \lambda a_2$ and $x^3\to\lambda^{-1}x^3$, 
the vector multiplet action is homogeneous of degree 1 and the 
chiral multiplet action is homogeneous of degree $-1$. This must be a symmetry 
of the action even when $\lambda$ is a function of $x^3$: the redefinition of $x^3$
commutes with $\cN=2$ 3d SUSY. We can write a Lagrangian where the coefficient 
of the kinetic term of the chiral multiplet is $1$, and the coefficient of the kinetic term 
of the gauge multiplet is $g^{-4}(x^3)$. This Lagrangian has no ambiguity. 
Then we can rescale by a factor of $\lambda = g^2(x^3)$, which 
restores the familiar normalization, but produces the correct ordering of the 
$D D_3 a_2$ term. If the coupling is 
introduced into \eqn{3d-adjoint} in this way, the last term is replaced, 
for a space-dependent coupling, by
\beq
g^2D\,D_3\frac{a_2}{g^2}\,.
\label{a2/g^2}
\eeq
Hence the kinetic term for the
scalars will take the form $\frac{1}{g^2} D_\mu a_1 D^\mu a_1 +
g^2 D_\mu \frac{a_2}{g^2} D^\mu \frac{a_2}{g^2}$. This is
familiar from the ${\cal N}=4$ SYM Janus domain wall \cite{D'Hoker:2006uv}.
We see that in the sharp interface limit, $g^{-2} a_2$ is the correct form of a scalar 
which is continuous across the interface. 

We would like to allow the 4d theta angle to vary as well.
A position dependent theta angle is not a total derivative anymore,
and affects the equations of motion. 
The gauge field boundary conditions are modified to 
the continuity of $\mathrm{Im} \tau F_{3i} + \mathrm{Re} \tau \frac{1}{2} \epsilon_{ijk} F^{jk}$ 
across the interface. Hence we expect to see boundary conditions for the scalars which 
impose the continuity, say, of $a_1$ and $a_1^D = \mathrm{Im} \tau a_2 + \mathrm{Re} \tau a_1$.

Upon integration by parts the theta angle term in the Lagrangian
becomes a Chern-Simons coupling in 3d. This can be completed to the 
$\cN=2$ CS Lagrangian
\begin{equation}
\frac{\partial_3 \theta}{8 \pi^2} \Tr\left(A \wedge F + 2 D a_1 \right).
\label{N=2CS}
\end{equation}
We integrate the first term back by parts to get the desired non-constant 4d 
$F\wedge F$ instanton term%
\footnote{The Janus wall involves the minimal modifications required for 
space-dependent couplings. The CS term will reappear in a dual description 
as a ``duality wall'' in Section~\ref{pestun}.}
but keep the supersymmetric completion, which modifies the scalar 
kinetic term. When we now integrate out the auxiliary field we get the term
\beq
g^2\left(D_3\frac{a_2}{g^2}+\frac{\partial_3\theta}{8\pi^2}a_1\right)^2
=g^2\left(\frac{1}{4\pi} D_3 a_1^D-\frac{\theta}{8\pi^2 } D_3 a_1\right)^2
\eeq
Here we introduced a new real field
\beq
a^D=a^D_1+ia^D_2=\tau a\,,
\qquad
\tau=\frac{\theta}{2\pi}+\frac{4\pi i}{g^2}\,.
\eeq
in this notation, for $\theta=0$, the field appearing in \eqn{a2/g^2} is 
$a^D_1$. Note that this kinetic term has a very natural form, invariant in the
abelian case under 
electro-magnetic dualities. Although we considered until now a free field
Lagrangian, it is natural to conjecture that this expression
will be correct for a more general IR prepotential as well, as
long as $a_1$ and $a^D_1$ are the real parts of the periods.
Notice that $a^D_1$ is the natural superpartner of the
restriction to 3d of the electro-magnetic dual gauge field.

For completeness we describe the possible modifications of the matter hypermultiplets Lagrangian. 
Each hypermultiplet
decomposes into two 3d chiral multiplets with scalars $q,\tilde q$ and 
auxiliary fields $F, \tilde F$. The 3d kinetic terms for each of them 
are as in \eqn{3d-chiral}. The 4d kinetic term is completed 
once we include the superpotential term 
$\Tr\tilde q (D_3 + i a_2) q$ which leads to the bosonic terms
\begin{equation}
\int dx^3\Tr\left(D_3 q D^3 \bar q + qa_2^2\bar q 
+ i q a_2\, D_3 \bar q +i (D_3 q) a_2\bar q 
+(q\to\tilde q)+ g^2 (q \tilde q)^2\right).
\label{superpot}
\end{equation}
This expression has some terms of the form 
$ia_2D_3(q\bar q+\tilde q\bar{\tilde q})$ which are not familiar 
in the 4d Lagrangian. New terms arise also when integrating out the $D$ 
auxiliary field. With the extra contributions from $q$ and $\tilde q$ 
(and no theta angle term), the D-term is
\beq
\frac{1}{g^2} D^2=
g^2\left(D_3\left( \frac{a_2}{g^2}\right) + q\bar q + \tilde q\bar{\tilde q}\right)^2.
\eeq
The mixed terms here combine with those form \eqn{superpot} to give
$g^2 D_3 \big(\frac{1}{g^2} a_2(q\bar q+\tilde q\bar{\tilde q})\big)$. 
In the sharp limit this can be integrated by parts to a term localized at the 
wall. Thus, in the presence of matter, the Janus domain wall includes 
extra terms localized at the wall. Notice that $a_2$ and the $q, \tilde q$
are set to zero by Pestun's localization, so this term does not affect the 
1-loop determinant. 

The hypermultiplets will generically have masses associated to flavor symmetries. 
In flat space they are complex masses, and can be interpreted as 
constant expectation values for a vectormultiplet scalar coupled to the flavor symmetry. 
They clearly break conformal invariance.
On $S^4$, it is also possible to turn on mass parameters in a similar fashion, 
but only real masses corresponding to constant expectation values of the real 
scalar in the vectormultiplet, $a_1$.

One may further wonder if extra cubic superpotential terms are
possible for the chiral multiplets which arise from the
hypermultiplets, which would be classically conformal invariant in 
flat space and could be carried over to $S^4$. 
The coefficient $C(x)$ of such a term should be a function which goes
to zero to the left and to the right of the wall, and should go to a $\delta$ function in the 
sharp limit, representing a superpotential integrated on the wall. On the other hand $C(x)^2$
enters the potential, so some care
is needed. The derivative of the overall superpotential with
respect to $\tilde q$ will be $\delta(x^3) \frac{\partial
W_3}{\partial \tilde q} + [D_3 + i a_2, q]$, and the $\delta$
function is eliminated by requiring a discontinuity of $q$ at
the sharp wall, proportional to $\frac{\partial W_3}{\partial
\tilde q}$. Upon localization, the discontinuity consistently vanishes, 
and it does not seem to affect the calculation at all. 

\subsection{Simple Boundary Conditions}
There are two natural boundary conditions for a gauge theory:
Dirichlet and Neumann. Dirichlet b.c. set $F_{ij}=0$ at the
boundary, and leave $F_{3i}$ free. Neumann boundary conditions
naively set $F_{3i}=0$ and leave $F_{ij}$ (and gauge
transformations) free at the wall. Actually. in the presence of
a theta angle, Neumann b.c. are modified to 
$F_{3i} + \frac{g^2\theta}{8\pi^2} \frac{1}{2} \epsilon_{ijk} F^{jk}=0$ 
\cite{Gaiotto:2008sa}. 
${\cal N}=2$ SUSY in 3d then extends the Dirichlet b.c. for the
gauge field to Dirichlet b.c. to $a_1$ and Neumann to $a_2$ or
$a_1^D$, while Neumann b.c. are extended to Dirichlet b.c. for
$a_1^D$ and Neumann for $a^1$.

This can be easily generalized to describe intermediate types
of boundary conditions or domain walls were the gauge symmetry
at the wall is reduced to a subgroup. Janus domain walls can be
seen as an example of this setup: a domain wall between two
identical gauge theories, possibly with different gauge
couplings, where the gauge symmetry is reduced to the diagonal
subgroup.

Consider a 4d theory on half of $S^4$ 
with Neumann boundary conditions at the equator.
For Neumann boundary conditions the value of $a_1$ is not fixed at the 
equator. Therefore, the partition function on half of $S^4$ 
will involve an integral over the Liouville momenta of a single conformal block 
multiplied by some function of the momenta from the one-loop contributions
\beq
\cZ_\text{Neumann}
=\int d\nu(\alpha) \bar g_\alpha\,\cF_\alpha(q)\,.
\eeq

\subsection{Coupling to 3d Matter} 
Whenever some gauge symmetry survives at the wall, one can
couple extra degrees of freedom there. In order to write down a
Lagrangian, all that is needed is to couple the 3d gauge
multiplet arising from the decomposition of the 4d vector
multiplet to the extra 3d matter theory. In particular, $a_1$
couples to the 3d theory as a mass term, and $a_2$ only through 
the $D_3 a_2$ term coupling to the 3d auxiliary fields. 

\section{S-Duality for ${\cal N}=4$ SYM}
\label{sec:S-duality}

A full understanding of general duality walls would
require us to know the detailed form of the 3d SCFT which lives
at the walls. We lack that information for general ${\cal
N}=2$ theories: if anything, the explicit form of the fusion
and braiding matrices could give suggestions about the nature
of such general 3d SCFTs. One relatively simple case is that of
a better understood theory: ${\cal N}=4$ SYM.

The duality wall which corresponds to the standard S-duality
operation in ${\cal N}=4$ SYM couples the theories on the 
two sides of the wall to a special 3d SCFT called $T[G]$.
This theory has ${\cal N}=4$ SUSY in 3d, a Higgs branch with a
$G$ action and a Coulomb branch with a $G^L$ action of the dual
group to $G$. The Higgs branch is coupled to the gauge theory
on one side of the wall, the Coulomb branch to the gauge theory
on the other side of the wall. The scalar fields $-ia^R$ and
$-ia^L$ on either sides of the wall act as mass parameters on
the respective branches. For generic values of $a^R$ and
$a^L$ the 3d theory is massive and all the off-diagonal modes
of the 4d theory are massive as well, leaving a $U(1)^r$
massless abelian gauge theory on both sides of the wall. The
massive 3d theory has several vacua, in correspondence with the
Weyl group of $G$. In each of the vacua, the abelian gauge field on the 
left and the right of the wall $A_L$ and $A_R$ can be explicitly 
dualized leading to an extra CS pairing $\langle A_L, w(A_R)\rangle$. 
Here $\langle\, ,\,\rangle$ is the Killing form, and $w$ the Weyl group element
associated with the 3d vacuum.

Using the results above we expect the partition function in the 
presence of the duality wall to take the form 
\beq
\exp\left[2 \pi i \left<w(a^L),a^R\right>\right].
\eeq
This is exactly the numerator in \eqn{FZZTbb} (with $m\to a^L$ 
and $a\to a^R$), which is proportional to the modual matrix 
\eqn{Sma}. 

So if $\cN=4$ SYM forces the 1-loop
determinants of all the other massive modes to cancel out, the $S^4$
partition function is
\beq 
\cZ =\int
d\nu(\alpha')d\nu(\alpha)\,{\overline
\cF^\text{torus}_{\alpha'}}(q')\,
S_{\alpha'\alpha}\,
\cF^\text{torus}_{\alpha}(q)\,.
\label{Z-Janus2} 
\eeq
This is indeed the modular matrix implementing the S-move on the torus.

\newpage
\end{fmffile}
\bibliography{refs}
\end{document}